%% file: Paper.tex
\documentclass[prb,twocolumn,showpacs,preprintnumbers,english,amsmath,amssymb]{revtex4}

\usepackage{graphicx,epsf}
\usepackage{dcolumn}
\usepackage{bm}
\input{macros}

\newcommand{\state}{coherence }

\renewcommand{\otimes}{\circ}
\newcommand{\bfm}[1]{\mbox{\boldmath $ #1 $}}

\newcommand{\sltilde}[1]{{\, \tilde{\! #1}}}
\newcommand{\slGamma}{{\it \Gamma}}
\newcommand{\slDelta}{{\it \Delta}}

\newcommand{\slPi}{{\it \Pi}}
\newcommand{\slSigma}{{\it \Sigma}}
\newcommand{\slPhi}{{\it \Phi}}
\newcommand{\slPsi}{{\it \Psi}}
\newcommand{\slOmega}{{\it \Omega}}

\begin{document}


\title{
The scattering problem in
non-equilibrium quasiclassical theory of metals
and superconductors:
general boundary conditions and applications }
\author{Matthias Eschrig} 
\affiliation{
Institut f\"ur Theoretische Festk\"orperphysik and 
DFG-Center for Functional Nanostructures,
Universit\"at Karlsruhe, D-76128 Karlsruhe, Germany}
\date{July 13, 2009}
\begin{abstract}
I derive a general set of boundary conditions for quasiclassical transport theory of 
metals and superconductors 
that is valid for equilibrium and non-equilibrium situations and includes
multi-band systems, weakly and strongly spin-polarized systems,
and disordered systems. 
The formulation is in terms of the normal state scattering matrix.
Various special cases for boundary conditions are known in the literature, that are
however limited to either equilibrium situations or single band systems.
The present formulation unifies and extends all these results.
In this paper I will present the general theory in terms of coherence functions and
distribution functions and demonstrate its use by applying it to the problem
of spin-active interfaces in superconducting devices and
the case of superconductor/half-metal interface scattering.
\end{abstract}
\maketitle

\section{Introduction}
\label{intro}
For the theoretical understanding of transport in metals and
superconductors Landau's concept of quasiparticles acting as elementary 
excitations over the ground state has been of immeasurable value.\cite{landau57,landau59}
In a normal metal, electrons are in a strongly quantum correlated state
due to Pauli's exclusion principle and due to Coulomb interactions.
Conduction electrons in metals are, however, quasiparticles, 
i.e. elementary excitations in the vicinity of the Fermi surface that are 
rarely scattering with each other as a result of
phase space restrictions. 
Although these quasiparticles are strongly coupled to electrons far away from 
the Fermi surface, renormalizations due to these interactions are constant over the
energy range of interest ($k_{\rm B}T$, with temperature $T$) and thus can be treated as
phenomenological parameters of the theory.\cite{landau57,landau59}
Quasiparticles are represented by a classical distribution function and obey   
a semiclassical Landau-Boltzmann transport equation.\cite{landau57}

Landau's Fermi liquid theory can be formulated in a systematic way within a
diagrammatic expansion of many-body Green's functions.\cite{footexp}
Asymptotic
expansion in the small parameter $k_{\rm B}T/E_{\rm F}$ (with the Fermi energy $E_{\rm F}$)
leads to the quasiclassical theory of metals,\cite{Eliashberg62,eliashberg71,serene83}
that describes the range $(k_{\rm B}T)^2/E_{\rm F}\ll k_{\rm B}T\ll E_{\rm F}$ in temperature well.
In leading order, the dynamical equations for Green's functions can 
be transformed into Landau's transport equation for quasiparticle distribution
functions.\cite{Abrikosov59,landau59,Eliashberg62,eliashberg71,serene83} 
Electrons that are far away from the Fermi surface and thus do no represent
quasiparticles enter this theory as effective interaction vertices.
Only a small number of these vertices is needed to describe the dynamics
of the quasiparticles.

The development of semiclassical concepts for the superconducting state 
was pioneered by \name{Geilikman}\cite{Geilikman58,Geilikman58a} and
\name{Bardeen}~{\em et al.}\cite{Bardeen59} soon after the development of
the BCS-theory of superconductivity.\cite{bardeen57}
Several early  works\cite{Larkin63,Ambegaokar64,Leggett65} on
transport and linear response  in superconductors
showed that various  semiclassical concepts of Landau's  Fermi liquid
theory could be readily generalized to  the superconducting 
state. A formulation of the equilibrium theory of superconductivity near 
the superconducting critical temperature $T_{\rm c}$
in terms of classical correlation functions was developed by
\name{de Gennes}.\cite{degennes66}

In the seminal works of
\name{Larkin} and \name{Ovchinnikov}\cite{larkin68} and
\name{Eilenberger}\cite{eilenberger} the concepts of
the BCS pairing theory of superconductors\cite{bardeen57} were
merged
with the concepts of Boltzmann transport equations
within Landau's Fermi liquid theory.
This quasiclassical theory of superconductivity was later
generalized to non-equilibrium phenomena by \name{Eliashberg}\cite{eliashberg71} 
and  \name{Larkin} and \name{Ovchinnikov}.\cite{larkin75} 

Quasiclassical methods can be applied to both
wave-function techniques and Green's function techniques.
In the former case, the starting point are
\name{Bogoljubov's} equations,\cite{Bogoliubov58,degennes66} leading
in quasiclassical approximation to Andreev's equations for
the envelopes of the waves.\cite{andreev64}
Alternatively, one can start from the microscopic
Nambu-Gor'kov matrix Green's functions.\cite{gorkov58}
In quasiclassical approximation they result into envelope
Green's functions that vary on
the coherence length scale, $\xi_0 =\hbar |\vf |/2\pi k_{\rm B} T_{\rm c}$
(with Fermi velocity $\vf $),
and the time scale $t_0=\hbar /\slDelta $ (with gap $\slDelta $), 
and are free of irrelevant fine-scale structures on the Fermi
wave length scale. 

Dynamical phenomena are described within quasiclassical theory
by using the Keldysh Green's function technique.\cite{keldysh64}
Quasiparticle states in superconductors 
are coherent mixtures of particle and hole states. The degree of mixing is
determined by the superconducting order parameter $\slDelta $.
The spectrum of quasiparticles is coupled to quasiparticle distribution functions,
and this coupling is expressed in Keldysh's technique
by two types of Green's functions, $g^{\ra }$ and $g^{\kel }$, that are elements
of a 2$\times $2 matrix $\check g$. The information about distribution functions is
in the Keldysh part, $g^{\kel }$.
Different formulations in terms of dynamical distribution functions in 
the superconducting state have been introduced
by Larkin and Ovchinnikov,\cite{larkin75} by Shelankov,\cite{shelankov85} and
by the author.\cite{eschrig97}

The derivation of boundary conditions for quasiclassical Green's functions is 
a difficult problem. For microscopic Green's functions the formulation of
boundary conditions, e.g. in terms of scattering matrices or
transfer matrices at interfaces, is rather simple. 
In contrast, in quasiclassical theory 
only the envelope function of the Bloch waves is known. 
The information about the phase 
of the waves is, however, missing.
Under these circumstances it is not a priori clear if boundary conditions
can be formulated within quasiclassical theory. That this is indeed
the case was shown independently by 
Shelankov,\cite{shelankov84} and by Zaitsev.\cite{zaitsev84}
More general formulations have been derived 
subsequently,\cite{ashauer86,kieselmann87,nagai88,millis88} including 
a formulation in terms of scattering matrices by Millis, Rainer and Sauls.\cite{millis88}
However, owing to the normalization condition for the quasiclassical propagator, the
boundary conditions so far were formulated as non-linear equations. Furthermore,
their practical use was limited as they contained unphysical, spurious solutions that 
lead to instabilities in numerical calculations.

Progress has been achieved by using the projector formalism 
of Shelankov,\cite{shelankov85} that allows an explicit formulation of boundary
conditions for both equilibrium \cite{yip97,eschrig00,ozana00}
and non-equilibrium\cite{eschrig00} situations. 
These boundary conditions have been generalized for
the single band case to include spin-active interfaces in 
equilibrium\cite{fogelstrom00} and in non-equilibrium,\cite{zhao04}
diffusive interface scattering,\cite{luck03} and
interfaces with strongly spin polarized ferromagnets.\cite{eschrig08,grein09}
An alternative, equivalent, route has been followed via
transfer matrices.\cite{cuevas01, huertas02,eschrig03,kopu04,graser07}
All the developments above were complemented by boundary conditions 
for diffusive superconductors\cite{kup88,nazarov99,berg08} that are 
appropriate for the diffusive limit of quasiclassical theory, 
the Usadel theory.\cite{usadel70}

In this work, we will pursue the approach in terms of scattering
matrices, and will present the boundary conditions in their most general form.
Our results include all previous formulations as special cases,
and are capable of describing e.g. non-equilibrium effects, multiband metals,
spin polarized systems, and diffusive interfaces.
In most of these cases the present formulation leads to more transparent and compact
boundary conditions, that allow (i) for a very effective numerical implementation
and (ii) better analytical treatment due to their simpler structure.
We use throughout the notation of Ref.~\onlinecite{eschrig00}.

\section{Theoretical description}
\label{theory}

Quasiclassical theory is a powerful tool for describing
inhomogeneous superconducting systems in and out of equilibrium, 
covering both ballistic and diffusive 
materials.\cite{larkin68,eilenberger,usadel70,schmid75,schmid81,serene83,rammer86,Larkin86,FLT} 
All the relevant physical information is contained in the
quasiclassical Green function
$\hat{g}(\epsilon, \pf , {\bfm R},t)$. Here
$\epsilon$ is the quasiparticle energy measured from the chemical
potential, $\pf $ the quasiparticle momentum on the Fermi
surface (that can have several branches),
${\bfm R}$ is the spatial coordinate, and $t$ is the time. The ``hat'' refers
to the 2$\times$2 matrix structure of the propagator in the Nambu-Gor'kov
particle-hole space, and the ``check'' to the 2$\times $2 Keldysh matrix structure.  
The equation of motion for $\hat{g}$ is the
Eilenberger equation,\cite{larkin68,eilenberger}
\beq
\label{eil}
\,[\ep \tc \check{1} - \check{h}, \check{g}]_{\qt } +i\hbar \qpartial \check{g}
= \check{0}
\eeq
subject to the normalization condition
\beq
\label{norm}
\check{g}\qt \check{g} = -\pi^2\, \check{1}.
\eeq
The elements of the 2$\times $2 Keldysh matrices are matrices in Nambu-Gor'kov 
particle-hole space,
\ber
\check{g} =
\left( \begin{array}{cc} \gmr & \gmk \\ 0 & \gma
\end{array} \right), \quad
\check{h} &=&
 \left( \begin{array}{cc}  \hr & \hk \\ 0 &  \ha
 \end{array} \right) ,
\end{eqnarray}
and $\hat \tau_3 $ is the third Pauli matrix in particle-hole space.
The $\qt $-product combines a time convolution and a 
matrix product and is explained in Appendix~\ref{Notation}.
For what follows it is useful to think about it as discretized in time, in which
case its properties are that of conventional matrix multiplication.\cite{footqt}
In equilibrium we will have to retain a matrix structure if 
the spin degree of freedom is active, in which case the $\qt $-product
reduces to a matrix multiplication in Pauli spin space.

Self energies enter Eq.~\eqref{eil} via the  matrices
\begin{eqnarray}
\hra &=&
 \left( \begin{array}{cc} \va & \Da \\ \Db & \vb
 \end{array} \right)^{\ra} ,\;
\hk =
\left( \begin{array}{cc} \plus \va & \plus \Da \\ -\Db & -\vb
\end{array} \right)^{\kel}  ,
\end{eqnarray}
where diagonal ($\va $) and off-diagonal ($\Da $) self energies are determined by
self-consistency equations. In this paper we do not, however, need to specify the 
exact form of these equations, and will assume for what follows that their
solutions are given.
There are fundamental symmetries that relate the particle and hole 
components of both self energies and Green's functions.\cite{serene83}
We express these symmetries throughout this paper by using the 
particle-hole conjugation operation that is defined in the mixed 
($\epsilon$, $t$) representation via
\beq
\tilde Q(z,\pf,\R,t)=Q(-z^\ast, -\pf,\R,t)^{\ast },
\eeq
where $z=\epsilon $ is real for the Keldysh components, and $z$ is situated in
the upper (lower) complex energy half plane for retarded (advanced) quantities.

The characteristic curves of the partial differential equation~\eqref{eil} define the
quasiclassical trajectories. Trajectories are labeled by the 
position on the Fermi surface, $\pf $, and are aligned with the Fermi velocity
$\vf (\pf) $. Quasiparticles move along these trajectories, thereby
being coherently coupled to the condensate.

Eqs.~\eqref{eil} and \eqref{norm} must be supplemented by boundary conditions
at the two ends of each trajectory. 
Eq.~\eqref{eil} is numerically stiff, with exponentially growing solutions
in both directions. In addition, unphysical solutions must be eliminated 
using the normalization condition Eq.~\eqref{norm}.
Both problems are solved in a natural way with
the parameterization of the quasiclassical Green's functions by
coherence and distribution functions.\cite{eschrig00}
These are physical quantities that obey initial value problems
with a stable integration direction, and automatically ensure the normalization
of $\check g$.

\subsection{Coherence functions and distribution functions}
\label{cohdist}
\begin{figure}[t]
\includegraphics[width = 1.0\columnwidth]{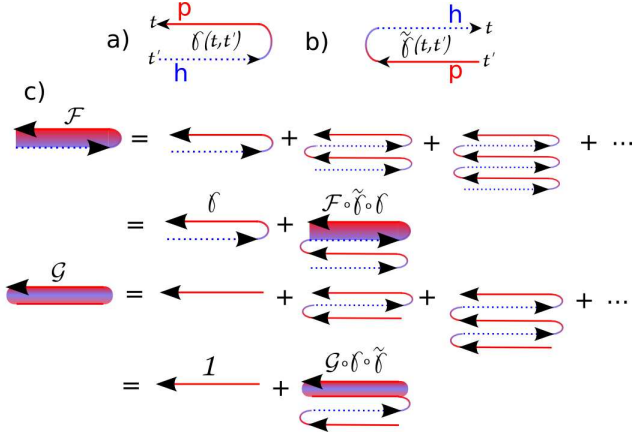}
\caption{\label{fig1}
(Color online)
a) The coherence function $\ga (t,t')$ describes the local probability 
amplitude for conversion of a hole (dotted line) at time $t'$ to a 
particle (full line) at time $t$.
For retarded functions $t>t'$, and for advanced functions $t<t'$.
b) The corresponding local amplitude for conversion of a particle at time $t'$ 
into a hole at time $t$
is described by the coherence function $\gb (t,t')$.
c) Diagrammatic representation of Eqs.~\eqref{FG0} and \eqref{FG}. 
}
\end{figure}
The numerical solution of 
the (non-linear) system of Eqs.~\eqref{eil} and \eqref{norm} is greatly simplified by
using a convenient parameterization of the 
Green functions in terms of retarded and advanced coherence functions $\gara $, $\gbra $,
and distribution functions $x$, $\tilde x$.\cite{eschrig97,eschrig99,eschrig00,eschrig04}
The coherence functions are a generalization of the so-called
Riccati amplitudes\cite{nagato93,schopohl95} 
to non-equilibrium situations.
Using a projector formalism as described in Appendices~\ref{appB} and
~\ref{appC} we can write the retarded and advanced Green's functions
[here the upper (lower) sign corresponds to retarded (advanced)] as
\ber
\label{cgretav}
\gqra 
&=& 
\mp \, 2\pi i\, 
\mat \plus {\cal G} & \plus {\cal F} \\ -\tilde{\cal F} &
-\tilde{\cal G} \matend^{\!\!\! \ra } 
\pm i\pi \hat \tau_3
, 
\eer
with the parameterization\cite{eschrig97}
\ber
\label{FG0}
{\cal G}&=&({\it 1}-\ga \qt \gb)^{-1}, \quad
{\cal F}=({\it 1}-\ga \qt \gb)^{-1}\qt \ga, \\
\label{tFG0}
\tilde{\cal G}&=&({\it 1}-\gb \qt \ga)^{-1}, \quad
\tilde{\cal F}=({\it 1}-\gb \qt \ga)^{-1}\qt \gb.
\eer
The inverse is defined via the $\otimes $-product, 
\beq
(\ldots )^{-1}\qt
(\ldots )= (\ldots )\qt(\ldots )^{-1}={\it 1},
\eeq
with the unit element ${\it 1}$ (see Appendix~\ref{Notation}).
Obviously, we can calculate the coherence functions from
\beq
\ga = {\cal G}^{-1}\qt {\cal F}= {\cal F}\qt \tilde{\cal G}^{-1}, \quad
\gb = \tilde{\cal G}^{-1}\qt \tilde{\cal F}= \tilde{\cal F}\qt {\cal G}^{-1}. 
\quad
\eeq
In order to obtain a diagrammatic representation we re-formulate
the problem in terms of Dyson equations
\ber
\label{FG}
{\cal G}&=& {\it 1} + {\cal G} \qt \ga \qt \gb , \quad
{\cal F}= \ga + {\cal F} \qt \gb \qt \ga , \\
\label{tFG}
\tilde{\cal G}&=& {\it 1} + \tilde{\cal G} \qt \gb \qt \ga ,\quad
\tilde{\cal F}= \gb + \tilde{\cal F}\qt \ga \qt \gb.
\eer
In Fig.~\ref{fig1} the corresponding diagrammatic expansion is shown.
Here, and in the following, 
we adopt and extend a diagrammatic notation by L\"ofwander, Zhao and 
Sauls.\cite{lofwander02,zhao07,zhao08}
The quantity ${\cal G}$ describes the local spectral 
amplitude of a particle-like excitation
in the presence of a condensate. This amplitude is renormalized
from its normal state value $\delta (t-t')$ due to multiple virtual Andreev scattering
processes that take place in the presence of an off-diagonal 
complex condensate field $\slDelta $.
The same holds for hole-like excitations, described by the quantity $\tilde{\cal G}$.
The ``anomalous'' propagators ${\cal F}$ and $\tilde{\cal F}$ result from
the local coherence amplitudes for particle-hole conversion,
$\gamma $, and for hole-particle conversion, $\tilde \gamma $, 
again by taking into account renormalization due to multiple virtual Andreev processes. 
For small superconducting amplitudes (e.g. near $T_{\rm c}$) the anomalous propagators
coincide with the coherence amplitudes.
\begin{figure}[b]
\includegraphics[width = 1.0\columnwidth]{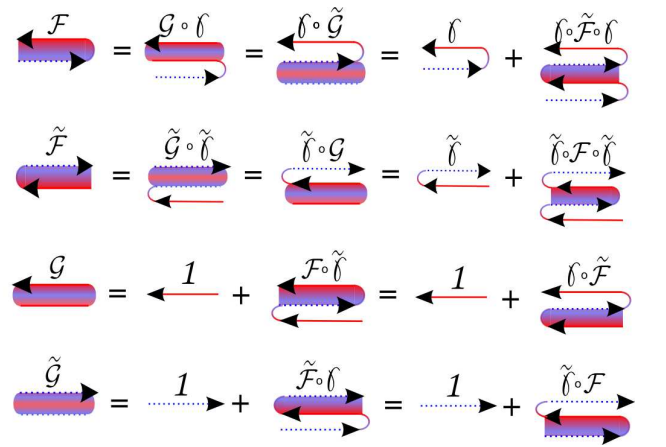}
\caption{\label{fig2}
(Color online)
Identities that hold between the six quantities ${\cal F}$, $\tilde{\cal F}$, ${\cal G}$,
$\tilde{\cal G}$, $\ga $, and $\gb $ as defined in Eqs.~\eqref{FG0}-\eqref{tFG0}.
}
\end{figure}
The four functions ${\cal F}$, $\tilde{\cal F}$, ${\cal G}$ and
$\tilde{\cal G}$ are
inter-related via $\ga $ and $\gb $, and a number of identities hold that are shown in 
Fig.~\ref{fig2} diagrammatically.

\begin{figure*}[t]
\includegraphics[width = 2.0\columnwidth]{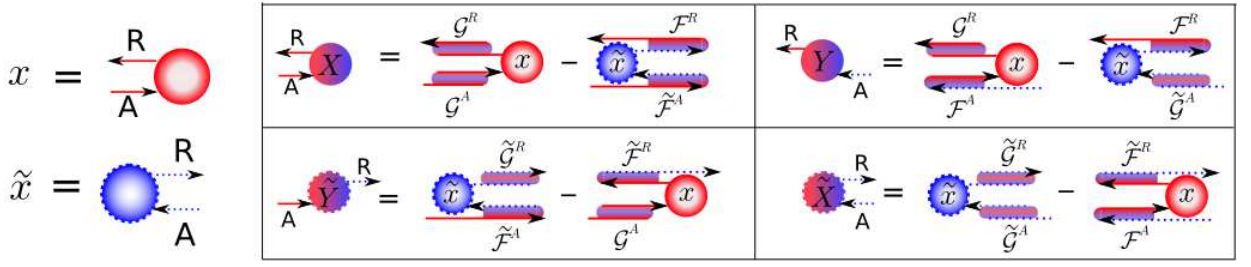}
\caption{\label{fig3}
(Color online)
The distribution functions $\xa$ and $\xb $ 
(left) connect incoming advanced and outgoing retarded propagators. 
The Keldysh components ${\cal X}^{\kel}$, $\tilde {\cal X}^{\kel}$,
${\cal Y}^{\kel}$, and $\tilde{\cal Y}^{\kel}$ are shown
in a diagrammatic representation 
of Eqs.~\eqref{Xkel}-\eqref{tYkel}.  ``${\rm R}$'' and ``${\rm A}$'' refer to 
``retarded'' and ``advanced''.
The particle distribution function $\xa $ and hole distribution function
$\xb $ are coherently mixed due to 
multiple coherent Andreev scattering events with
amplitudes given by the renormalized quantities
${\cal G}$, $\tilde {\cal G}$, ${\cal F}$, and $\tilde {\cal F}$ that are
sums of terms shown in Fig.~\ref{fig1}. 
}
\end{figure*}
Although the coherence functions $\ga $ and $\gb$ are sufficient to describe
the retarded and advanced Green's functions, 
the quantities in Eqs.~\eqref{FG}-\eqref{tFG}
allow for an effective formulation of boundary
conditions (see below).
The Keldysh part of the propagator can be formulated in
terms of these and a suitable distribution function for particle-like and 
hole-like excitations, respectively, in the following way,
\ber
\label{ckelgf2}
&&\gqk \equiv
-2\pi i  
\mat \plus {\cal X} & \plus {\cal Y} \\ \tilde{\cal Y} &
\tilde{\cal X} \matend^{\!\!\! \kel }  \\
&&= -2\pi i  
\mat \plus {\cal G} & \plus {\cal F} \\ -\tilde{\cal F} &
-\tilde{\cal G} \matend^{\!\!\! \ret } \qt
\mat \xa & 0 \\ 0 & \xb \matend \qt
\mat \plus {\cal G} & \plus {\cal F} \\ -\tilde{\cal F} &
-\tilde{\cal G} \matend^{\!\!\! \adv } .
\nonumber 
\eer
Making use of the identities in Fig.~\ref{fig2},
we can further use that ${\cal X}^{\kel }=
{\cal G}^{\ret} \qt\xa \qt {\cal G}^{\adv}- {\cal F}^{\ret} \qt \xb \qt \tilde{
\cal F}^{\adv} =
{\cal G}^{\ret} \qt (\xa - \gar \qt \xb \qt \gba ) \qt {\cal G}^{\adv}
$ and similarly for the other components, 
which gives
\ber
\label{Xkel}
{\cal X}^{\kel }&=&
{\cal G}^{\ret} \qt (\xa - \gar \qt \xb \qt \gba ) \qt {\cal G}^{\adv}
, \\
{\cal Y}^{\kel }
&=& {\cal G}^{\ret}\qt (\xa \qt \ga^{\adv} -\ga^{\ret} \qt \xb )\qt \tilde{\cal G}^{\adv} , 
\\ 
\tilde{\cal X}^{\kel }&=&
\tilde{\cal G}^{\ret} \qt ( \xb - \gb^{\ret} \qt \xa \qt \ga^{\adv} ) \qt \tilde{\cal G}^{\adv} ,
\\
\tilde{\cal Y}^{\kel }&= &
\tilde{\cal G}^{\ret} \qt (\xb \qt \gb^{\adv} -\gb^{\ret} \qt \xa )\qt {\cal G}^{\adv} .
\label{tYkel}
\eer
The Keldysh amplitudes ${\cal X}$, $\tilde{\cal X}$, ${\cal Y}$, and $\tilde{\cal Y}$ are
shown in a diagrammatic representation in Fig.~\ref{fig3}.
Note that for the Keldysh components we need to keep track of retarded and
advanced coherence functions. As advanced functions propagate backward in
time, their group velocity is reversed. Advanced propagators can be described
as usual by the particle-antiparticle paradigm, that in the present case
is equivalent to a particle-hole transformation
as described in Appendix~\ref{symm}. In drawing diagrams we prefer
to keep the particle picture instead of introducing
antiparticles (which would reverse the arrows and turn them into hole propagators with opposite energy).

We stress that there are no diagrams
with more than one $\xa $ or $\xb $ vertex, as no retarded propagator can
enter either of them, and no advanced propagator can emerge from them. 
As a result,
the structure of the equations for ${\cal X} $, $\tilde{\cal X} $, ${\cal Y}$,
and $\tilde {\cal Y}$ formally corresponds to that of a linear response with a 
perturbation that switches from retarded to advanced (in fact, the 
linear response theory for retarded and advanced coherence functions
has many formal similarities with the Keldysh part of the
transport theory,\cite{eschrig99} see also Appendix~\ref{properties}).

\subsubsection{Alternative distribution functions}

Other
definitions for distribution functions have been introduced in the literature.
We discuss the issue of the various possibilities in defining distribution
functions and their relation with each other in detail in Appendix~\ref{distpar}.
The distribution functions $h$ introduced by Larkin and Ovchinnikov
\cite{larkin68,Larkin86} and $F$ introduced by 
Shelankov \cite{shelankov85} are related to the distribution functions
$\xa $ and $\xb $ by
\ber
\label{xaF}
\xa &=& \Fa - \gar \qt \Fa \qt \gba = 
h + \gar \qt \tilde h \qt \gba 
\; , \nonumber \\
\xb &=& \Fb - \gbr \qt \Fb \qt \gaa = \tilde h + \gbr \qt h \qt \gaa 
.
\eer
Series expansions for the
inverses can be obtained by iteration,\cite{cuevas06} for example
\ber
\mbox{$\Fa= \sum_{n=0}^\infty (\gar )^n \qt \xa \qt (\gba )^n $}
\eer
with $(\ldots)^n = (\ldots)^{n-1}\qt (\ldots)$, and
\ber
\label{xtoh}
\mbox{$
h= \sum_{n=0}^\infty (\gar \gbr )^n \qt (\xa -\gar \qt \xb \qt \gba ) \qt (\gaa \gba )^n.
$}
\eer
In equilibrium,
\beq
h_{\rm eq}=\Fa_{\rm eq}=\tanh \frac{\ep}{2T}=-\Fb_{\rm eq} = -\tilde h_{\rm eq}
\eeq
holds.
The advantages of the functions $\xa $ and $\xb $ are that the
transport equations take their simplest form,\cite{eschrig97}
their numerical evaluation is easier, they simplify considerably
time-dependent problems,\cite{eschrig97,eschrig99,cuevas06}
and as we will show below,
they allow for an effective handling of the boundary conditions.

\subsection{Transport equations}

The central equations that govern the transport phenomena have been derived
in Ref.~\cite{eschrig97,eschrig99,eschrig00}.
The transport equation for the coherence functions
$\gamma (\epsilon, {\bfm p}_{\rm F}, {\bfm R},t)$  and
$\tilde \gamma (\epsilon, {\bfm p}_{\rm F}, {\bfm R},t)$
are given by
\ber
\label{cricc1}
&& 
(i\hbar \, \qpartial +2\ep )\gara = [ \ga \qt \Db \qt \ga + 
 \va \qt \ga - \ga \qt \vb  - \Da ]^{\ra}, \nonumber \\
\label{cricc2}
&& 
(i\hbar \, \qpartial -2\ep ) \gbra = [ \gb \qt \Da \qt \gb + 
 \vb \qt \gb - \gb \qt \va   - \Db ]^{\ra} .\nonumber \\
\label{cricc}
\eer
For the distribution functions 
$x(\epsilon, {\bfm p}_{\rm F}, {\bfm R},t)$  and
$\tilde x(\epsilon, {\bfm p}_{\rm F}, {\bfm R},t)$
the transport equations read
\ber
\label{keld1}
&&(i\hbar \, \qpartial + i\hbar \, \partial_t ) \xa -[\ga \qt \Db +\va ]^{\ret } \qt \xa -
\xa \qt [ \Da \qt \gb -\va ]^{\adv }  \nonumber \\
&&\qquad =-\gar \qt \vbk \qt \gba + \Dak \qt \gba + \gar \qt \Dbk  -\vak ,
\\
\label{keld2}
&&(i\hbar \, \qpartial - i\hbar \, \partial_t) \xb -[ \gb \qt \Da +\vb ]^{\ret} \qt \xb -
\xb \qt [ \Db \qt \ga - \vb ]^{\adv}  \nonumber \\
&&\qquad =-\gbr \qt \vak \qt \gaa +\Dbk \qt \gaa + \gbr \qt \Dak - \vbk .
\eer
In Appendix~\ref{properties} we discuss properties of the solutions of
these equations, and equivalent formulations in terms of integral equations.

An important property of the set of equations \eqref{cricc1}-\eqref{keld2} is 
their invariance with respect to gauge transformations.
There are two types of gauge transformations that are important, and that are
very different in nature. We discuss this issue in Appendix~\ref{gauge}.
The first type is the usual gauge invariance that links the phase of the
coherence functions with the electromagnetic potentials. The second
type leaves retarded and advanced quantities invariant and affects
only the distribution functions $\xa $ and $\xb $ and the Keldysh part of
the self energies. It leads to a certain freedom for the choice of the
distribution functions (several choices have been mentioned above). 
In particular, when a reference system is present, distribution
functions can be defined with respect to those
of the reference system. They are then called ``anomalous'',\cite{footanom}
and vanish in
the reference system. This is particularly useful for situations 
when a system is coupled to a reservoir.

\subsubsection{Homogeneous equilibrium solution}

In the case that both 
${\cal E}^{\ra} = \epsilon - (\vara-\vbra)/2$ and $[\Da \Db]^\ra$ are
diagonal in spin space, 
the homogeneous solutions for the \state functions
in equilibrium can be written as,
\ber
\label{hom1}
\gara_{\rm h,eq} = -\left[\frac{\Da }{{\cal E} \pm i\sqrt{-\Da \Db - {\cal E}^2 }} 
\right]^{\ra }
\eer
where the upper (lower) sign holds for the retarded (advanced) functions. 
For a singlet superconductor in the clean limit $[\Da \Db]^{\ra } = -|\slDelta|^2$.
In the presence of a constant superflow with momentum $\vec{p}_s$ one has to replace
$\epsilon $ by $\epsilon - \vf \cdot \vec{p}_s$.

For the distribution function in equilibrium one obtains
\beq
\xa_{\rm h,eq} = 
(1-\gar \gba )
\tanh(\frac{\epsilon }{2T})   .
\eeq
Note that $\gba = (\gar )^\dagger $ (see Appendix~\ref{symm}).

\subsubsection{General solution for homogeneous self energies}

For homogeneous self energies we can express the solutions $\gara (\rho)$ 
along a certain trajectory with path variable $\rho $ (defined by the
trajectory parameterization $\vec{R}=\vec{R}_0+\rho \vec{v}_{\rm F}$),
for a given initial value $\gara (0) =\gara_0$,
in terms of the homogeneous solution $\gara_{\rm h} $ that satisfies
\ber
\label{ghom}
&& 
[ \ga_{\rm h} \qt \Db \qt \ga_{\rm h} - \Ea \qt \ga_{\rm h} + \ga_{\rm h} \qt \Eb  - \Da ]^{\ra}=0,
\eer
where $\Eara = \epsilon - \vara $, $\Ebra = -\epsilon - \vbra$.
Defining $\slOmega^{\ra}_1=[\Ea-\ga_{\rm h} \qt \Db ]^\ra$ and
$\slOmega^{\ra}_2=[\Eb+\Db \qt \ga_{\rm h} ]^\ra $, and using the relations of
Appendix~\ref{relations}, it follows as
\ber
\gara(\rho )=\left[\ga_{\rm h} + 
e^{i\rho \slOmega_1} \qt
\delta_0 \qt \big\{e^{i\rho \slOmega_2}+ C (\rho)\qt \delta_0\big\}^{-1}\right]^\ra,
\eer
with
$\delta^\ra_0= \left[\ga_0-\ga_{\rm h}\right]^\ra $ and
\ber
C^\ra(\rho )&=& \left[C_0\qt e^{i\rho \slOmega_1}-e^{i\rho \slOmega_2} \qt C_0 
\right]^\ra
\eer
where $C_0^\ra $ is the solution of the equation
\beq
[C_0\qt \slOmega_1 - \slOmega_2 \qt C_0]^\ra = \Dbra  .
\eeq
For equilibrium we have $\Eara = -\Ebra \equiv {\cal E}^\ra$, and
if ${\cal E}^{\ra} $ and $[\Da \Db]^\ra$ are diagonal in spin space, 
then with $\slOmega^\ra_1=-\slOmega^\ra_2\equiv \slOmega^\ra $
the relation
\beq
\gara(\rho) = \left[\frac{\ga_0 \slOmega + i\tan(\rho \slOmega ) (\Ea \ga_0 + \Da )}{\slOmega - i\tan(\rho \slOmega ) (\Ea-\ga_0 \Db )} \right]^\ra 
\eeq
follows, in agreement with Ref.~\onlinecite{kalenkov07}.

\subsubsection{Equilibrium solution for sub-gap energies in the 
presence of an inhomogeneous order parameter in the clean limit}
\label{gpsi}

If we can neglect impurity scattering, and the system outside the
scattering region is asymptotically homogeneous with gap $\slDelta_{\rm h}$,
then for sub-gap energies $|\epsilon |\le |\slDelta_{\rm h}|$ we can make some more
general statments about the properties of the coherence amplitudes.
In particular, if we e.g. consider a pure singlet pairing state, and
if the order parameter is of the form $\slDelta=\slDelta_0(\rho ) e^{i\chi }i\sigma_y$ with
spatially varying modulus $\slDelta_0$ and spatially constant phase $\chi $,
then, using the ansatz $\gar (\rho )= i\ga_0^{\,}(\rho) e^{i[\chi+\slPsi(\rho )] } \cdot i\sigma_y$ with real $\ga_0^{\,}$ and $\slPsi $,
the equilibrium equations of motion along any fixed trajectory read
\ber
\label{geq1}
\frac{{\rm d}\gamma_0^{\,}}{{\rm d}\rho } &= & \plus (1-\gamma_0^2)\slDelta_0 \cos (\slPsi) 
-0^+ \gamma_0^{\,}
\\
\gamma_0^{\,}\frac{{\rm d}\slPsi }{{\rm d}\rho}&=& -(1+\gamma_0^2)\slDelta_0 \sin (\slPsi )
+2 \epsilon \gamma_0^{\,},
\eer
where $0^+$ is a positive infinitesimal.
The first equation is stable only in direction of increasing $\rho $.
Now, for the initial condition far away from the scatterer, for sub-gap energies
$|\epsilon | \le |\slDelta_{\rm h}|$ the relation $\gamma_0^{\,}=1$ holds.
Then, as Eq.~\ref{geq1} shows, 
this property will be preserved along the trajectory
regardless of the spatial variation of $\slDelta_0(\rho)$.
That means, only the phase $\slPsi $ of the coherence amplitude varies, and we have
$\gar = i e^{i[\chi+\slPsi(\rho ) ]} \cdot i\sigma_y$ with
\ber
\label{psi}
\frac{{\rm d}\slPsi (\rho )}{{\rm d}\rho}&=& -2\slDelta_0 (\rho) \sin \big(\slPsi (\rho)\big) +2 \epsilon 
\eer
and initial condition $\slPsi(\rho_0)=0$.
If $\epsilon=0$, then the coherence amplitude stays constant along the trajectory.
Similarly, we obtain
$\gbr = -i e^{-i[\chi+\sltilde\slPsi(\rho ) ]} \cdot i\sigma_y$ with
\ber
\frac{{\rm d}\sltilde\slPsi (\rho )}{{\rm d}\rho}&=& 2\slDelta_0 (\rho) \sin \big(\sltilde\slPsi (\rho)\big) +2 \epsilon .
\eer
For energies $|\epsilon|> |\slDelta_{\rm h}|$ both the modulus and phase of $\gar $, $\gbr $ vary
in space.
A similar consideration can be made for any unitary order parameter.

\section{Scattering Theory}

We consider in the following a general quantum mechanical scattering problem that is 
characterized by incoming and outgoing Bloch wave solutions. 
We assume that the scattering region is localized in a certain space area, where
we have in mind e.g. an interface, a surface, or an impurity. 
In quasiclassical context there will be trajectories
that enter and leave the scattering region.
Correspondingly we can define incoming solutions along each trajectory
as those for which
the group velocity is pointing towards the scattering region 
under consideration (the ``scatterer''), 
and outgoing those for which the group velocity is pointing away. 
The projection of the group velocity on the Fermi momentum has one and the same
sign for $\gar $, $\gba $, and $\xa $, and the opposite sign for
$\gbr $, $\gaa $, and $\xb $. Correspondingly, these six objects
for each trajectory
always group into three incoming and three outgoing ones.

The scatterer will lead to a mixing between the trajectories that enter
the scattering region. Depending on symmetry constraints, the possible 
number of scattering wave vectors might be drastically reduced,
as for example is the case for conservation of parallel momentum 
at an atomically clean interface. 
In the latter case, for a single Fermi surface on each side
of the interface, there will be mixing
only between the incoming, reflected, transmitted trajectory, and a fourth trajectory
that is reached by a process involving ``crossed'' Andreev reflection. 
In the case of
a diffusive interface trajectories of all directions will be mixed with each
other.

In order to distinguish incoming and outgoing directions we will adopt the 
notation of Ref.~\onlinecite{eschrig00}, that small case letters
$\gara $, $\gbra $, $\xa $, $\xb $ denote incoming quantities, and
capital case letters $\Gara$, $\Gbra$, $\Xa $, $\Xb$ denote outgoing
quantities. 
As the quasiclassical Green's function is parameterized by the momentum
$\pf $, it is composed of both incoming and outgoing quantities.
We may write the Keldysh matrix Green's function as a functional of
the four coherence functions and the two distribution functions.
If the Fermi velocity points towards the scatterer, this functional
dependence will be
\beq
\check{g}=\check{g}[\gar ,\Gbr, \Gaa, \gba, \xa, \Xb ],
\eeq
and for the case that the Fermi velocity points away from the scatterer,
it is
\beq
\check{g}=\check{g}[\Gar ,\gbr, \gaa, \Gba, \Xa, \xb ].
\eeq

Usually the potentials in a scattering region vary on a energy scale large
compared to the superconducting gap or the temperature. In this case, 
it is sufficient to know the {\it normal state scattering matrices} 
for particle-like excitations, denoted by $\vec{S} $ with elements
$S(\pf\leftarrow \pf' )$, and for hole-like
excitations, denoted by $\mbf{\tilde{S}}$ with elements $\tilde S(\pf' \leftarrow \pf )$, 
that connect outgoing with incoming
quasiparticles on trajectories parameterized by the Fermi momenta 
$\pf $ and $\pf' $.\cite{footscatt}
The scattering matrix in particle-hole space reads
\begin{eqnarray}
\hat {\bfm S}=
\left( \begin{array}{cc} 
{\bfm S} & 0\\0 & 
{\tilde{\bfm S}}^\dagger
\end{array} \right)   
\qquad
\hat{\bfm S}^\dagger =
\left( \begin{array}{cc} 
{\bfm S}^\dagger  & 0\\0 & 
{\tilde{\bfm S}}
\end{array} \right)   .
\end{eqnarray}
In order to reduce the amount of notation
we will in the following label trajectories with
the Fermi velocity pointing away from the scatterer
simply by $k$, $k'$, $k_1$ etc,
and trajectories with
the Fermi velocity pointing towards the scatterer
by $p$, $p'$, $p_1$ etc, thus 
omitting the vector notation. It is understood that those labels 
are from the set of Fermi momenta associated with all the trajectories
that overlap with the scattering region.
As for the discussion in this chapter the dynamical variables (energy, time)
enter only as parameters, we will suppress the dependence on these.
We assume for the scattering problem that the spatial coordinate 
$\vec{R}$ on each trajectory entering or leaving the scattering region
is sufficiently far from the scatterer in order that the scattered waves have 
taken their asymptotic form, but sufficiently close to neglect spatial
variations on the scale of the coherence length in the scattering region, 
and we will suppress these spatial coordinates in this chapter as well.
The scattering problem will thus be fully characterized
by the set of $k$ and $p$ values associated with all involved trajectories.
In a centro-symmetric system
(or a non-centrosymmetric system with time reversal symmetry), 
for each $k$ value there will also be the 
trajectory with the opposite direction $p=-k$.

It is our task to express the set of outgoing coherence and
distribution functions $\Gar_k $,$\Gbr_p $, 
$\Gaa_p $,$\Gba_k $, $\Xa_k $ $\Xb_p $ by the incoming
ones $\gar_p $,$\gbr_k $, 
$\gaa_k $,$\gba_p $, 
$\xa_p $ $\xb_k $ for a given scattering matrix
$S_{kp}$ (the scattering matrix $\tilde{S}_{pk}$ for hole-like excitations 
is related to that for particle-like excitations by the particle-hole conjugation symmetry).

\subsection{Elementary interface Andreev scattering events}

The central objects for the formulation of boundary conditions for the
coherence functions and distribution functions are the following
quantities, that express an elementary scattering event,
\begin{eqnarray}
\label{gp}
[\gamma'_{kk'}]^{\ret } &=& [\sum_{p} S_{kp} \qt \gamma_{p} \qt \tilde S_{pk'}]^{\ret } ,\\
\label{gpa}
[\gamma'_{pp'}]^{\adv } &=& [\sum_{k} S_{pk} \qt \gamma_{k} \qt \tilde S_{kp'}]^{\adv} ,\\
\label{xp}
x'_{kk'} &=& \sum_{p} S^{\ret}_{kp} \qt x_{p} \qt S^{\adv}_{pk'},
\end{eqnarray}
together with the respective particle-hole conjugated quantities,
\begin{eqnarray}
\label{tgp}
[\tilde \gamma'_{pp'}]^{\ret } &=& [\sum_{k} \tilde S_{pk} \qt \tilde \gamma_{k} \qt S_{kp'}]^{\ret}\\
\label{tgpa}
[\tilde \gamma'_{kk'}]^{\adv } &=& [\sum_{p} \tilde S_{kp} \qt \tilde \gamma_{p} \qt S_{pk'}]^{\adv}\\
\label{txp}
\tilde x'_{pp'} &=& \sum_{k} \tilde S^{\ret}_{pk} \qt \tilde x_{k} \qt \tilde S^{\adv}_{kp'}.
\end{eqnarray}
As we will show below, the scattering matrices enter the boundary conditions only
in terms of these quantities. This allows for a compact matrix
notation. For example, we can re-formulate boundary conditions
for spin active interfaces that are known in literature,\cite{fogelstrom00,zhao04} 
in a rather compact way. Importantly, a straightforward generalization
of these boundary conditions to multiple bands, to disordered interfaces, 
to strongly spin polarized ferromagnets, to strongly spin-orbit split bands, 
and to the general scattering problem from a target is possible.
For equilibrium we recover also the results by Shelankov and Ozana,\cite{ozana00} 
that were obtained by a similar procedure.
In order to switch to a compact matrix notation, we 
introduce the diagonal matrices 
\ber
\gamma_{kk'}^{\ret}&=&\gamma_k^{\ret} \, \delta_{kk'}, \quad
\gamma_{pp'}^{\adv}=\gamma_p^{\adv} \, \delta_{pp'}, \quad
x_{kk'}=x_k \, \delta_{kk'}, \qquad \\
\tilde \gamma_{pp'}^{\ret}&=&\tilde \gamma_p^{\ret} \, \delta_{pp'}, \quad
\tilde \gamma_{kk'}^{\adv}=\tilde \gamma_k^{\adv} \, \delta_{kk'}, \quad
\tilde x_{pp'}=\tilde x_p \, \delta_{pp'}.
\eer
With this, we can write the elementary scattering events as
\begin{eqnarray}
\label{gp1}
[\mbf{\gamma}']^{\ra} &=& [\mbf{S}\qt \mbf{\gamma} \qt \mbf{\tilde S}]^{\ra},
\mbf{x}' = \mbf{S}^{\ret} \qt \mbf{x} \qt \mbf{S}^{\adv}, \qquad\\
\label{gpt1}
[\mbf{\tilde \gamma}']^{\ra} &=& [\mbf{\tilde S} \qt \mbf{\tilde \gamma} \qt \mbf{S}]^{\ra}, 
\mbf{\tilde x}' = \mbf{\tilde S}^{\ret} \qt \mbf{\tilde x} \qt \mbf{\tilde S}^{\adv}.
\qquad
\end{eqnarray}
\begin{figure}[t]
\includegraphics[width = 1.0\columnwidth]{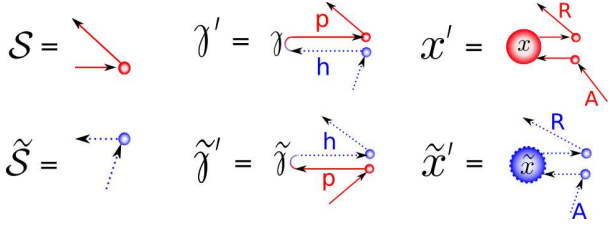}
\caption{\label{fig4}
(Color online)
Diagrammatic symbols for the elementary scattering events described by 
Eqs.~\eqref{gp1}-\eqref{gpt1}. $p$ and $h$ refers to ``particle'' and ``hole'', and ``${\rm R}$'' and ``${\rm A}$'' to ``retarded'' and ``advanced''. A sum over internal
variables according to Eqs.~\eqref{gp}-\eqref{xp} is implied.
}
\end{figure}
In Fig.~\ref{fig4} we show these scattering events in diagrammatic form.
We note that the retarded and advanced scattering matrices are related
by fundamental symmetry,
\beq
\mbf{S}^{\adv} = [\mbf{S}^{\ret}]^\dagger ,\quad
\mbf{\tilde S}^{\adv} = [\mbf{\tilde S}^{\ret}]^\dagger ,
\eeq
which leads together with the symmetries in Appendix~\ref{symm} to the
symmetry relations
\ber
\,[\mbf{\gamma}']^{\adv } &=& [\mbf{\tilde\gamma}']^{\ret \dagger} ,\quad
\mbf{x}'= [\mbf{x}']^{\dagger } ,\\
\,[\mbf{\tilde \gamma}']^{\adv } &=& [\mbf{\gamma}']^{\ret \dagger} ,\quad
\mbf{\tilde x}' = [\mbf{\tilde x }']^{\dagger }.
\eer

\subsection{Derivation of boundary conditions}

\subsubsection{Retarded Propagator}

\begin{figure}[t]
\includegraphics[width = 1.0\columnwidth]{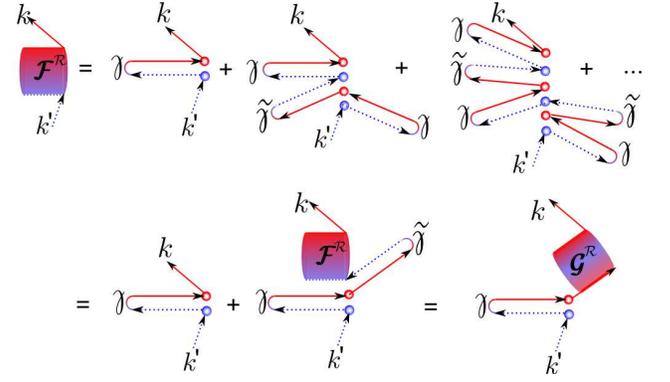}
\caption{\label{fig5}
(Color online)
Diagrammatic representation of 
Eq.~\eqref{gpp} for the retarded functions. 
In the last line the identity~\eqref{GtoF} is shown diagrammatically,
which defines the diagrammatic expansion for ${\cal G}^{\ret}$.
Summation over internal variables is implied.
}
\end{figure}

The anomalous functions ${\cal F}^{\ret}$ are obtained from a sum over all
virtual multiple Andreev scattering events that are {accompanied by
interface scattering}. We consider first the set of retarded Green's functions
with directions $k$ that are directed away from the scatterer.
In this case, the retarded propagator is given by
\beq
\hat{g}_k^{\ret }=\hat{g}^{\ret}_k[\Gar ,\gbr ].
\eeq
We introduce effective interface coherence
amplitudes as solutions of the equation
\begin{eqnarray}
\label{gpp}
\, [{\cal F}_{kk'}]^{\ret } &=& \left[ \gamma'_{kk'} 
+\sum_{k_1} 
{\cal F}_{kk_1} \qt \tilde \gamma_{k_1} \qt \gamma'_{k_1k'}
\right]^{\ret } . 
\end{eqnarray}
Using a compact matrix notation, the solutions are
\begin{eqnarray}
\label{gpp2}
\, \mbf{\cal F}^{\ret } &=& \left[ 
\mbf{\gamma}'  \qt \left( \mbf{1}-\mbf{\tilde \gamma} \qt \mbf{\gamma}' \right)^{-1}
\right]^{\ret } , 
\end{eqnarray}
where the inversion $\mbf{Q}^{-1} $ is defined via
$\mbf{Q} \qt \mbf{Q}^{-1} =\mbf{1}$ with $1_{kk'}=\delta_{kk'}{\it 1} $.
The diagrammatic representation of this expansion is
shown in Fig.~\ref{fig5}. 
We also define the corresponding particle-hole diagonal interface amplitude 
\begin{eqnarray}
\label{gpp3}
\, \mbf{\cal G}^{\ret} &=& \left[ 
\left( \mbf{1}-\mbf{\gamma}'  \qt \mbf{\tilde \gamma} \right)^{-1}
\right]^{\ret}
= 1+ \left[ \mbf{\cal F} \qt \mbf{\tilde \gamma} \right]^{\ret }, 
\end{eqnarray}
that is closely related to the function $\mbf{\cal F}^{\ret } $ by 
\ber
\label{GtoF}
\, \mbf{\cal F}^{\ret} &=& \left[ \mbf{\cal G} \qt \mbf{\gamma}' \right]^{\ret }. 
\eer
For the quasiclassical approximation only the component $k'=k$ is relevant,
as it contributes to the slowly varying envelope function of trajectory $k$, 
and we obtain
\ber
{\cal G}^{\ret}_k &=& {\cal G}^{\ret}_{kk}, \quad
{\cal F}^{\ret}_k = {\cal F}^{\ret}_{kk}  \quad .
\eer
The remaining two retarded Green's function matrix components are $\tilde{\cal G}^{\ret}_k=[(
\mbf{1}- \mbf{\tilde \gamma} \qt \mbf{\gamma}'  )^{-1} ]^{\ret}_{kk}$ and
$ \tilde {\cal F}^{\ret}_k = \tilde{\cal G}^{\ret }_{k} \qt \gbr_k $. 
According to section~\ref{cohdist},
for the outgoing coherence functions the equation
${\cal F}^\ret_{kk}=\Gar_k+{\cal F}^\ret_{kk} \qt \gbr_k\qt \Gar_k =
({\it 1}+{\cal F}^\ret_{kk} \qt \gbr_k )\qt \Gar_k $
holds, which according to
Eq.~\ref{gpp3} is equal to ${\cal G}^\ret_{kk}\qt \Gar_k$.
Thus, we extract the outgoing coherence amplitudes from the solution of the equation
\ber
\label{Eqret}
{\cal G}^{\ret}_{kk} \qt \Gar_{k\leftarrow k'} &=& {\cal F}^{\ret}_{kk'}, \quad
\Gar_k=\Gar_{k\leftarrow k}  \quad .
\eer
The more general quantity $\Gar_{k\leftarrow k'}$ that is introduced here
will be needed below, e.g. in the transport equations for the distribution functions.

For the component $\Gbr $ we must consider the retarded Green's function 
for a direction $p$ towards the scatterer, as the
group velocity of $\Gbr $ is opposite to the direction of the momentum.
The corresponding retarded propagator is given by,
\beq
\hat{g}^{\ret}_p=\hat{g}^{\ret}_p[\gar ,\Gbr ].
\eeq
We obtain in complete analogy to the discussion above
\begin{eqnarray}
\label{tgpp2}
\, \mbf{\tilde{\cal F}}^{\ret } &=& \left[ 
\mbf{\tilde \gamma}'  \qt \left( \mbf{1}-\mbf{\gamma }\qt \mbf{\tilde \gamma}' \right)^{-1}
\right]^{\ret } = 
\left[ \mbf{\tilde{\cal G}} \qt \mbf{\tilde \gamma}' \right]^{\ret } ,\quad \\
\, \mbf{\tilde{\cal G}}^{\ret} &=& \left[ 
\left( \mbf{1}-\mbf{\tilde \gamma}' \qt \mbf{\gamma } \right)^{-1}
\right]^{\ret} 
=1+ \left[ \mbf{\tilde{\cal F}} \qt \mbf{ \gamma} \right]^{\ret } ,
\end{eqnarray}
from which we extract the outgoing coherence amplitude
by solving the equation
\ber
\label{tEqret}
\tilde{\cal G}^{\ret}_{pp} \qt \tilde\Ga^{\ret}_{p\leftarrow p'} &=& \tilde{\cal F}^{\ret}_{pp'}, \quad
\tilde\Ga^{\ret}_p=\tilde\Ga^{\ret}_{p\leftarrow p}  \quad .
\eer

\subsubsection{Advanced Propagator}

For the advanced functions we need to take into account that they propagate backward in
time. Thus, we consider for $\Gaa $ the advanced Green's function for a 
direction $p$ towards the scatterer, 
\beq
\hat{g}^{\adv}_p=\hat{g}^{\adv}_p[\Gaa, \gba ],
\eeq
and for $\Gba $ for a direction $k$ away from the scatterer,
\beq
\hat{g}^{\adv}_k=\hat{g}^{\adv}_k[\gaa, \Gba ].
\eeq
The most convenient form of the corresponding equations is
\begin{eqnarray}
\label{tgpp3}
\, \mbf{\cal F}^{\adv } &=& \left[ 
\mbf{\gamma}'  \qt \left( \mbf{1}-\mbf{\tilde\gamma }\qt \mbf{\gamma}' \right)^{-1}
\right]^{\adv } = 
\left[ \mbf{\gamma}' \qt \mbf{\tilde{\cal G}} \right]^{\adv } ,\quad \\
\, \mbf{\tilde{\cal G}}^{\adv} &=& \left[ 
\left( \mbf{1}-\mbf{\tilde \gamma} \qt \mbf{\gamma}' \right)^{-1}
\right]^{\adv} 
=1+ \left[ \mbf{\tilde \gamma} \qt \mbf{\cal F} \right]^{\adv }, 
\end{eqnarray}
with the the coherence amplitudes
\ber
\label{Eqadv}
\Ga^{\adv}_{p'\rightarrow p} \qt \tilde{\cal G}^{\adv}_{pp} 
&=& {\cal F}^{\adv}_{p'p}, \quad
\Ga^{\adv}_p=\Ga^{\adv}_{p\rightarrow p}  \quad ,
\eer
and
\begin{eqnarray}
\label{tgpp4}
\, \mbf{\tilde{\cal F}}^{\adv } &=& \left[ 
\mbf{\tilde \gamma}'  \qt \left( \mbf{1}-\mbf{\gamma }\qt \mbf{\tilde \gamma}' \right)^{-1}
\right]^{\adv } = 
\left[ \mbf{\tilde \gamma}' \qt \mbf{\cal G} \right]^{\adv } ,\\
\, \mbf{\cal G}^{\adv} &=& \left[ 
\left( \mbf{1}-\mbf{\gamma }\qt \mbf{\tilde \gamma}' \right)^{-1}
\right]^{\adv} 
=1+ \left[ \mbf{ \gamma} \qt \mbf{\tilde{\cal F}} \right]^{\adv } ,
\end{eqnarray}
with the the coherence amplitudes
\ber
\label{tEqadv}
\Gb^{\adv}_{k'\rightarrow k} \qt {\cal G}^{\adv}_{kk} 
&=& \tilde{\cal F}^{\adv}_{k'k}, \quad
\Gb^{\adv}_k=\Gb^{\adv}_{k\rightarrow k}  \quad .
\eer

\subsubsection{Keldysh Propagator}

\begin{figure}[t]
\centerline{
\includegraphics[width = 0.8\columnwidth]{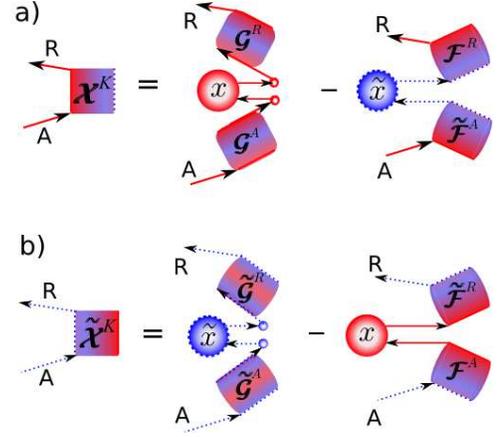}
}
\caption{\label{fig6}
(Color online)
Diagrammatic representation of Eqs.~\eqref{XXXa} and \eqref{tXXXa}, using
the identities in Eq.~\eqref{GtoF} and \eqref{tgpp2}.
}
\end{figure}

The corresponding expressions for the 
Keldysh components are obtained in a similar way. 
We perform a diagrammatic expansion
of the Keldysh components in the elementary scattering events,
using the fact that the vertices $\xa $ and $\xb $ can only occur once
in each diagram (see end of section \ref{cohdist}).
Thus, all renormalizations affect only the particle-hole
conversion processes on either side of the $\xa $ and $\xb $ vertices.
We consider first the Keldysh Green's function 
for $k$ being directed away from the scatterer, 
\beq
\hat{g}^{\kel}_k=\hat{g}^{\kel}_k[\Gar ,\gbr, \gaa, \Gba, \Xa, \xb ],
\eeq
for which the expansion, shown in Fig.~\ref{fig6}a, gives
\ber
\label{XXXa}
\mbf{\cal X}^{\kel }&=&
\mbf{\cal G}^{\ret} \qt (\mbf{\xa}' - {\mbf{\ga}'}^{\ret} \qt \mbf{\xb} \qt {\mbf{\gb}'}^{\adv} ) \qt \mbf{\cal G}^{\adv} . 
\eer
We obtain the distribution functions $X$ 
in terms of $\mbf{\cal X}^{\kel }$ by
\begin{eqnarray}
\label{XX2}
&&
{\cal G}_{kk}^{\ret} \qt \left(
X_{k} -\slGamma_{k}^{\ret} \qt \tilde x_{k} \qt \sltilde\slGamma_{k}^{\adv}
\right) \qt
\tilde{\cal G}_{kk}^{\adv} 
=  
{\cal X}^{\kel }_{kk} \quad .
\end{eqnarray}
Similarly, considering the Keldysh
Green's function 
for $p$ being directed towards the scatterer, 
\beq
\hat{g}^{\kel}_p=\hat{g}^{\kel}_p[\gar ,\Gbr, \Gaa, \gba, \xa, \Xb ],
\eeq
we obtain from the expansion shown in Fig.~\ref{fig6}b
\ber
\label{tXXXa}
\mbf{\tilde{\cal X}}^{\kel }&=&
\mbf{\tilde{\cal G}}^{\ret} \qt 
(\mbf{\xb}' - {\mbf{\gb}'}^{\ret} \qt \mbf{\xa} \qt {\mbf{\ga }'}^{\adv})
\qt \mbf{\tilde{\cal G}}^{\adv} 
,
\eer
and from this $\tilde X$ in terms of $\mbf{\tilde{\cal X}}^{\kel }$,
\begin{eqnarray}
\label{XXt2}
&&
\tilde{\cal G}_{pp}^{\ret} \qt \left(
\tilde X_{p} 
-\sltilde\slGamma_{p}^{\ret} \qt x_{p} \qt \slGamma_{p}^{\adv}\right) 
\qt {\cal G}_{pp}^{\adv} 
= 
{\tilde{\cal X}}^{\kel }_{pp}  \quad .
\end{eqnarray}

\subsubsection{Boundary conditions for coherence amplitudes}

For the outgoing coherence amplitudes that where obtained in Eqs.
\eqref{Eqret}, \eqref{tEqret},
\eqref{Eqadv}, and \eqref{tEqadv}, closed equations can be derived, that
can again be represented diagrammatically in a straightforward way.
We can cast 
$\slGamma_{k\leftarrow k'}^{\ret} =
({\cal G}_{kk}^{\ret} )^{-1}\qt {\cal F}_{kk'}^{\ret} $,
$\sltilde \slGamma_{p\leftarrow p'}^{\ret} = 
(\tilde{\cal G}_{pp}^{\ret} )^{-1}\qt
\tilde{\cal F}_{pp'}^{\ret} $,
$\slGamma_{p'\rightarrow p}^{\adv} =
{\cal F}_{p'p}^{\adv} \qt ({\cal G}_{pp}^{\adv })^{-1} $, and
$\sltilde \slGamma_{k'\rightarrow k}^{\adv} = 
\tilde{\cal F}_{k'k}^{\adv} \qt
(\tilde{\cal G}_{kk}^{\adv})^{-1} $,
in the form of Dyson-type equations,
\begin{eqnarray}
\label{GR}
\, [\slGamma_{k\leftarrow k'}]^{\ret} &=& \big[ \gamma'_{kk'} 
+\sum_{k_1\ne k} 
\slGamma_{k\leftarrow k_1} \qt \tilde \gamma_{k_1} \qt \gamma'_{k_1k'}
\big]^{\ret} \\
\label{tGR}
\, [\sltilde{\slGamma }_{p\leftarrow p'}]^{\ret} &=& \big[ \tilde{\gamma }'_{pp'} 
+\sum_{p_1\ne p} 
\sltilde \slGamma_{p\leftarrow p_1} \qt \gamma_{p_1} \qt \tilde \gamma'_{p_1p'}
\big]^{\ret} 
\end{eqnarray}
and
\begin{eqnarray}
\label{GA}
[\slGamma_{p'\rightarrow p}]^{\adv} &=& \big[ \gamma'_{p'p} 
+\sum_{p_1\ne p} 
\gamma'_{p'p_1 } \qt \tilde \gamma_{p_1} \qt
\slGamma_{p_1\rightarrow p} 
\big]^{\adv}\\
\label{tGA}
\,[\sltilde \slGamma_{k'\rightarrow k}]^{\adv} &=& \big[ \tilde \gamma'_{k'k} 
+\sum_{k_1\ne k} 
\tilde \gamma'_{k'k_1 } \qt \gamma_{k_1} \qt
\sltilde \slGamma_{k_1\rightarrow k} 
\big]^{\adv} \; .
\end{eqnarray}
From those we obtain the quasiclassical coherence amplitudes,
\begin{eqnarray}
\label{GRA}
\slGamma_{k}^{\ret }&=& \slGamma_{k \leftarrow k}^{\ret} \quad ,\qquad
\slGamma_{p}^{\adv }= \slGamma_{p \rightarrow p}^{\adv} \quad ,\\
\label{tGRA}
\sltilde \slGamma_{p}^{\ret }&=& \sltilde \slGamma_{p \leftarrow p}^{\ret} \quad ,\qquad
\sltilde \slGamma_{k}^{\adv }= \sltilde \slGamma_{k \rightarrow k}^{\adv } \quad .
\end{eqnarray}
The diagrammatic representation of these equations is the same as for the functions 
${\cal F}^{\ra }$ and $\tilde{\cal F}^{\ra }$, with the modification that in all
internal sums the direction $k$ of the final state that is scattered into is excluded.

\subsubsection{Boundary conditions for distribution functions}

Analogously to the discussion for the coherence amplitudes we derive now the
boundary conditions for the distribution functions. For this we
formally solve Eqs.~\eqref{XX2} and \eqref{XXt2},
\begin{eqnarray}
\label{X2}
&&X_{k} -\slGamma_{k}^{\ret} \qt \tilde x_{k} \qt \slGamma_{k}^{\adv}=  
[({\cal G}_{kk})^{-1}]^{\ret} \qt
{\cal X}_{kk} \qt
[(\tilde{\cal G}_{kk})^{-1}]^{\adv}
\\
\label{Xt2}
&&\tilde X_{p} 
-\sltilde\slGamma_{p}^{\ret} \qt x_{p} \qt \sltilde \slGamma_{p}^{\adv}= 
[(\tilde{\cal G}_{pp})^{-1}]^{\ret} \qt
{\tilde{\cal X}}_{pp} \qt
[({\cal G}_{pp})^{-1}]^{\adv} \qquad
\end{eqnarray}
and use the relations 
\ber
\, [({\cal G}_{kk})^{-1} \qt {\cal G}_{kk'}]^{\ret} &=& 1- [\overline \slGamma_{k\leftarrow k'}\qt \tilde \gamma_{k'}]^{\ret},  \\
\, [{\cal G}_{p'p} \qt ({\cal G}_{pp})^{-1}]^{\adv} &=& 1-[\tilde \gamma_{p'} \qt \overline \slGamma_{p' \rightarrow p}]^{\adv}, 
\eer
and the corresponding particle-hole conjugated equations. 
In these equations we have introduced the
scattering parts of the coherence functions, that is obtained by subtracting
the forward scattering contributions,
\begin{eqnarray}
\label{GbarR}
\,[\overline \slGamma_{k \leftarrow k'}]^{\ret } &=&
[\slGamma_{k \leftarrow k'} - \slGamma_{k} \delta_{kk'} ]^{\ret} \\
\label{GbarA}
\,[\overline \slGamma_{p' \rightarrow p}]^{\adv } &=&
[\slGamma_{p' \rightarrow p} - \slGamma_{p} \delta_{pp'} ]^{\adv }  \quad 
\end{eqnarray}
and similarly for particle-hole conjugated quantities.
Solving Eqs.~\eqref{X2} and \eqref{Xt2} for 
$X_k$ and $\tilde X_p$ leads to an explicit solution
in terms of $x'_k$, $\tilde x_k$, $\tilde x'_p$, and $x_p$,
\begin{widetext}
\begin{eqnarray}
\label{X}
X_{k} &= &
\sum_{k_1,k_2} 
[\delta_{kk_1}  + \overline \slGamma_{k \leftarrow k_1} \qt \tilde \gamma_{k_1}]^{\ret}
\qt x'_{k_1k_2} \qt
[\delta_{k_2k}  + \gamma_{k_2}\qt \overline{\sltilde \slGamma}_{k_2 \rightarrow k} ]^{\adv}
-\sum_{k_1}
[\overline \slGamma_{k \leftarrow k_1}]^{\ret} \qt \tilde x_{k_1} \qt
[\overline{\sltilde \slGamma}_{k_1 \rightarrow k} ]^{\adv}
\\
\label{tX}
\tilde X_{p} &= &
\sum_{p_1,p_2} 
[\delta_{pp_1}  + \overline {\sltilde \slGamma}_{p \leftarrow p_1} \qt \gamma_{p_1}]^{\ret}
\qt \tilde x'_{p_1p_2} \qt
[\delta_{p_2p}  + \tilde \gamma_{p_2}\qt \overline{\slGamma}_{p_2 \rightarrow p} ]^{\adv}
-\sum_{p_1}
[\overline {\sltilde \slGamma}_{p \leftarrow p_1}]^{\ret} \qt x_{p_1} \qt
[\overline{\slGamma}_{p_1 \rightarrow p} ]^{\adv}
\end{eqnarray}
\end{widetext}
The diagrammatic representation of these equations is the same as for the functions 
${\cal X}^{\ra }$ and $\tilde{\cal X}^{\ra }$, with the modification that in all
internal sums over virtual particle-hole or hole-particle conversion processes
the direction $k$ of the state that is scattered into is excluded.
The scattering into the final state (forward scattering) is taking place only in the
last scattering event. Note that these simple diagrammatic rules 
result from our particular choice of the distribution functions.
Applying a gauge transformation of the type discussed in
Appendix~\ref{gauge2} to the distribution functions amounts to shifting terms between
the two contributions on the right hand sides in Fig.~\ref{fig6} back and forth.
This leads to redefined distribution functions without changing the Keldysh Green's 
function.

\subsubsection{General use of boundary conditions}

Equations \eqref{gp}-\eqref{txp}, \eqref{GR}-\eqref{tGRA} and \eqref{GbarR}-\eqref{tX} give the outgoing quantities 
$\slGamma_k^{\ret}$, $\sltilde \slGamma_p^{\ret}$, 
$\slGamma_p^{\adv}$, $\sltilde \slGamma_k^{\adv}$, 
$X_k$, $\tilde X_p$ 
in terms of the incoming quantities 
$\gamma_p^{\ret}$, $\tilde \gamma_k^{\ret}$, 
$\gamma_k^{\adv}$, $\tilde \gamma_p^{\adv}$, 
$x_p$, $\tilde x_k$,
and are the main result of this paper.
For a small number of trajectories involved in the scattering process these
equations can be solved analytically.
For numerical calculations, in particular when many trajectories
are involved that mix with each other in the scattering region (diffusive
scattering), it might
be of advantage to use matrix algebra and solve the set of equations
\eqref{gp1}-\eqref{gpt1}, 
\eqref{gpp2}-\eqref{Eqret}, 
\eqref{tgpp2}-\eqref{tEqret}, 
\eqref{tgpp3}-\eqref{tEqadv}, 
\eqref{XXXa}-\eqref{XX2}, and
\eqref{tXXXa}-\eqref{XXt2}. The solution of these equations involves
only standard numerical linear algebra and is straightforward.

\section{Application I: Spin-active interface scattering in superconducting devices}
\label{sectappl2}

In this section we show how to recover from our formulation of
boundary conditions the results of
Refs.~\onlinecite{eschrig00}, \onlinecite{fogelstrom00}, and \onlinecite{zhao04}.
These boundary conditions are for an interface between two superconductors
or two metals or one superconductor and one metal. On both sides of the
interface each trajectory is doubly degenerate due to the spin degree of
freedom. The interface is assumed to conserve the momentum component 
parallel to the interface, $\vec{p}_{||}$. It is assumed that only 
one Fermi surface sheet is present in each material, such that only
one incoming and one outgoing trajectory exists for each side of the 
interface. For such a case the boundary conditions take a particular
simple form. 
As on either side of the interface (index 1 and 2) only one incoming and
one outgoing momentum direction are coupled by the boundary condition,
we can label the involved trajectories simply by indices 1 and 2, and
incoming and outgoing components 
by small and capital letters in the boundary condition.

We start with writing down Eqs.~\eqref{gp}-\eqref{xp} for this case:
\ber
\label{g11}
&&\left( \begin{array}{cc}  \ga'_{11}&  \ga'_{12}\\ \ga'_{21} & \ga'_{22}
\end{array} \right)^{\!\!\ra} = \nonumber \\
&&\qquad \left[
\left( \begin{array}{cc}  S_{11}&  S_{12}\\ S_{21} & S_{22}
\end{array} \right) \qt
\left( \begin{array}{cc}  \ga_1&  0\\ 0 & \ga_2
\end{array} \right) \qt
\left( \begin{array}{cc}  \tilde S_{11}&  \tilde S_{12}\\ \tilde S_{21} & \tilde S_{22}
\end{array} \right)\right]^{\ra} \qquad
\eer
and
\ber
\label{x11}
&&\left( \begin{array}{cc}  x'_{11}&  x'_{12}\\ x'_{21} & x'_{22}
\end{array} \right) = \nonumber \\
&&\qquad 
\left( \begin{array}{cc}  S_{11}&  S_{12}\\ S_{21} & S_{22}
\end{array} \right)^{\!\!\ret } \qt
\left( \begin{array}{cc}  x_1&  0\\ 0 & x_2
\end{array} \right) \qt
\left( \begin{array}{cc}  S_{11}&  S_{12}\\ S_{21} & S_{22}
\end{array} \right)^{\!\!\adv } \qquad
\eer
where all involved quantities are 2$\times $2 spin matrices.

\subsection{Coherence functions}

Using these quantities, the boundary condition Eq.~\eqref{GR} 
takes on the form
\ber
\label{ga11ret}
\,[\Ga_{1\leftarrow 1}]^{\ret} &=& [ \ga'_{11} + \Ga_{1\leftarrow 2} \qt \gb_2 \qt \ga'_{21} ]^{\ret} \\
\label{ga12ret}
\,[\Ga_{1\leftarrow 2}]^{\ret} &=& [ \ga'_{12} + \Ga_{1\leftarrow 2} \qt \gb_2 \qt \ga'_{22} ]^{\ret} \\
\label{ga21ret}
\,[\Ga_{2\leftarrow 1}]^{\ret} &=& [ \ga'_{21} + \Ga_{2\leftarrow 1} \qt \gb_1 \qt \ga'_{11} ]^{\ret} \\
\label{ga22ret}
\,[\Ga_{2\leftarrow 2}]^{\ret} &=& [ \ga'_{22} + \Ga_{2\leftarrow 1} \qt \gb_1 \qt \ga'_{12} ]^{\ret} .
\eer
The equations for the 12- and 21-components, 
Eqs.~\eqref{ga12ret} and \eqref{ga21ret},
can be solved directly,
\ber
\label{Gret12}
\,[\Ga_{1\leftarrow 2}]^{\ret} &=& [ \ga'_{12} \qt (1 - \gb_2 \qt \ga'_{22})^{-1} ]^{\ret} \\
\label{Gret21}
\,[\Ga_{2\leftarrow 1}]^{\ret} &=& [ \ga'_{21} \qt (1-  \gb_1 \qt \ga'_{11} )^{-1}]^{\ret}.
\eer
Analogously, for the advanced components Eq.~\eqref{GA} leads to
\ber
\label{Gadv12}
\,[\Ga_{1\rightarrow 2}]^{\adv} &=& [ (1- \ga'_{11}\qt \gb_1 )^{-1}\qt \ga'_{12} ]^{\adv} \\
\label{Gadv21}
\,[\Ga_{2\rightarrow 1}]^{\adv} &=& [ (1- \ga'_{22} \qt \gb_2)^{-1} \qt \ga'_{21} ]^{\adv} .
\eer
Introducing these into the corresponding 11- and 22-components, i.e.
Eqs.~\eqref{ga11ret} and \eqref{ga22ret} and analogously for the advanced 
functions,
gives the first set
of boundary conditions for the coherence functions,
\ber
\label{BCinterf1}
\,[\Ga_{1}]^{\ra} &=& [ \ga'_{11} + \ga'_{12} \qt (1 - \gb_2 \qt \ga'_{22})^{-1}\qt \gb_2 \qt \ga'_{21} ]^{\ra} \\
\label{BCinterf2}
\,[\Ga_{2}]^{\ra} &=& [ \ga'_{22} + \ga'_{21} \qt (1 - \gb_1 \qt \ga'_{11})^{-1}\qt \gb_1 \qt \ga'_{12} ]^{\ra} .\qquad
\eer
The particle-hole conjugated equations are obtained by simply applying the
particle-hole conjugation operation to these results.
These boundary conditions, together with the 
definitions~\eqref{g11},
are equivalent to the boundary conditions of
Ref.~\onlinecite{fogelstrom00}, and for spin-scalar scattering matrices to
those of Ref.~\onlinecite{eschrig00}.

\subsection{Distribution functions}

Turning to the the Keldysh components, we formulate
Eq.~\ref{X} for our case,
\ber
X_1&=&x'_{11}+\Gar_{1\leftarrow 2}\qt \gbr_2\qt x'_{21} + x'_{12} \qt \gaa_2 \qt \Gba_{2\rightarrow 1} 
\nonumber \\
&&\qquad \qquad + \Gar_{1\leftarrow 2}\qt (\gbr_2 \qt x'_{22}\qt \gaa_2-\tilde x_2 ) \qt \Gba_{2\rightarrow 1} \qquad \\
X_2&=&x'_{22}+\Gar_{2\leftarrow 1}\qt \gbr_1 \qt x'_{12} + x'_{21} \qt \gaa_1\qt \Gba_{1\rightarrow 2} 
\nonumber \\
&&\qquad \qquad + \Gar_{2\leftarrow 1}\qt (\gbr_1 \qt x'_{11}\qt \gaa_1 -\tilde x_1 ) \qt \Gba_{1\rightarrow 2} \qquad
\eer
Substituting Eqs.~\eqref{Gret12}-\eqref{Gadv21} into these gives the 
required boundary conditions for the distribution functions.
Again, the particle-hole conjugated equations are obtained by simply applying the
particle-hole conjugation operation to these results.
These boundary conditions, together with the 
definitions~\eqref{x11},
are equivalent to the ones of Ref.~\onlinecite{zhao04}, and for spin-scalar 
scattering matrices to those of Ref.~\onlinecite{eschrig00}.

\subsection{Spin-active interface in bilayer geometry}
\label{bilayer}

As an application we discuss the coherence functions for 
a bilayer that consists of a thick superconducting
layer (that we will treat as bulk system) with a thin normal metal overlayer
of thickness $d$. We consider a spin-active interface with a scattering matrix
\ber
S= \left( \begin{array}{cc}  r_{\rm S}&  t_{\rm SN}\\ t_{\rm NS} & -r_{\rm N} 
\end{array} \right)
= \tilde S^\ast .
\eer
We assume that the interface has a unique quantization axis,
in which case all reflection 
($r_{{\rm S}\uparrow }$, $r_{{\rm S}\downarrow}$, $r_{{\rm N}\uparrow }$, $r_{{\rm N}\downarrow}$)
and transmission amplitudes 
($t_{{\rm SN}\uparrow }$, $t_{{\rm SN}\downarrow}$, $t_{{\rm NS}\uparrow }$, $t_{{\rm NS}\downarrow}$) 
are spin diagonal.
We consider a singlet superconductor with (retarded) coherence amplitudes
$\gar = \ga_{\rm S} i\sigma_y$. 
As a result, all possible
induced correlations in the normal metal are written as
$\gar = \mbox{diag}[\ga_{{\rm N}\uparrow},\ga_{{\rm N}\downarrow}] i\sigma_y$, where
`diag' denotes a diagonal spin matrix with the diagonal elements as indicated.
In the following we restrict our discussion to the equilibrium situation.

Eqs.~\eqref{BCinterf1}-\eqref{BCinterf2} result into
\ber
&&\Ga_{{\rm N}\uparrow} =
\Big[r_{{\rm N}\uparrow} r^\ast_{{\rm N}\downarrow} \ga_{{\rm N}\uparrow} +
t_{{\rm N}{\rm S}\uparrow} t^\ast_{{\rm S}{\rm N}\downarrow} \ga_{\rm S} + 
\Big. \nonumber \\ \Big.
&&+\gb_{\rm {\rm S}}\ga_{\rm {\rm S}} \ga_{{\rm N}\uparrow} 
(r_{{\rm N}\uparrow} r_{{\rm S}\uparrow} + t_{{\rm N}{\rm S}\uparrow} t_{{\rm S}{\rm N}\uparrow})
(r^\ast_{{\rm N}\downarrow} r^\ast_{{\rm S}\downarrow} + t^\ast_{{\rm N}{\rm S}\downarrow} t^\ast_{{\rm S}{\rm N}\downarrow}) \Big] /
\nonumber \\
&&/ \Big[
1+\gb_{\rm S}( t_{{\rm S}{\rm N}\uparrow} t^\ast_{{\rm N}{\rm S}\downarrow} \ga_{{\rm N},\uparrow } +
r_{{\rm S}\uparrow} r^\ast_{{\rm S}\downarrow} \ga_{\rm S})\Big]
\eer
Now, using the unitarity condition of the scattering matrix, we write
(with $\sigma=\{\uparrow,\downarrow\}$),
$r_{{\rm N}\sigma} = r_{\sigma}e^{i\vartheta_{{\rm N}\sigma}}$,
$r_{{\rm S}\sigma} = r_{\sigma}e^{i\vartheta_{{\rm S}\sigma}}$,
$t_{{\rm N}{\rm S}\sigma} = t_{\sigma}e^{i\vartheta_{{\rm N}{\rm S}\sigma}}$,
$t_{{\rm S}{\rm N}\sigma} = t_{\sigma}e^{i\vartheta_{{\rm S}{\rm N}\sigma}}$, where
$\vartheta_{{\rm S}\sigma}+\vartheta_{{\rm N}\sigma}= \vartheta_{{\rm S}{\rm N}\sigma}+\vartheta_{{\rm N}{\rm S}\sigma}$,
and $r_{\sigma}^2+t_{\sigma}^2=1$.
Then, with the spin mixing angles 
$\vartheta_{{\rm S}}=\vartheta_{{\rm S}\uparrow}-\vartheta_{{\rm S}\downarrow} $ and
$\vartheta_{{\rm N}}=\vartheta_{{\rm N}\uparrow}-\vartheta_{{\rm N}\downarrow} $,
and with the further abbreviations 
$[(\vartheta_{{\rm S}{\rm N}\uparrow} +\vartheta_{{\rm S}{\rm N}\downarrow})
-(\vartheta_{{\rm N}{\rm S}\uparrow} +\vartheta_{{\rm N}{\rm S}\downarrow})]/2 = \vartheta'$,
$(\vartheta_{\rm {\rm N}}+\vartheta_{\rm {\rm S}})/2=\vartheta_+$, 
$(\vartheta_{\rm {\rm N}}-\vartheta_{\rm {\rm S}})/2=\vartheta_-$, 
the last equation becomes
\ber
\label{GaN}
\Ga_{{\rm N}\uparrow} 
\Big[e^{-i\vartheta_+}
+
\ga_{{\rm N}\uparrow } \gb_{\rm S} e^{i\vartheta'} t_{\uparrow}t_{\downarrow} 
+ \ga_{\rm S} \gb_{\rm S} e^{-i\vartheta_-} r_{\uparrow}r_{\downarrow} 
\Big] = && \nonumber \\ 
\ga_{{\rm N}\uparrow}
e^{i\vartheta_-} r_{\uparrow}r_{\downarrow}+
\ga_{\rm S} e^{-i\vartheta'} t_{\uparrow}t_{\downarrow} +
\ga_{{\rm N}\uparrow} \ga_{\rm S}\gb_{\rm S} e^{i\vartheta_+}
&& .
\eer
The extra spin-scalar phase $\vartheta'$ may appear due to 
time reversal symmetry breaking by the interface.
In order to obtain the coherence amplitudes at the
outer surface of the normal layer, $\gamma_{{\rm B}\uparrow}$,
we solve the transport equation
$(i\hbar \vec{v}_{\rm F} \cdot \nabla +2\epsilon )\ga_{\uparrow}(x) =0$ in the normal metal
with perfect reflection at the outer boundary,
which gives
\ber
\label{prop}
\gamma_{{\rm B}\uparrow}=\Ga_{{\rm N}\uparrow}e^{iz/\varepsilon_d}, \quad
\ga_{{\rm N}\uparrow}=\gamma_{{\rm B}\uparrow} e^{iz/\varepsilon_d},
\eer
with $z=\epsilon +i0^+$, $\varepsilon_d= \hbar v_{{\rm F}x}/2d$ the
ballistic Thouless energy, and $v_{{\rm F}x }$ the Fermi velocity component normal to 
the interface in the normal conductor.
We now concentrate on sub-gap energies, $|\epsilon| < |\slDelta |$.
Substituting Eq.~\eqref{prop} into Eq.~\eqref{GaN},
and using the bulk solutions
$\ga_{\rm S}=ie^{i\slPsi } =-\gb $ with $\slPsi=\arcsin(\epsilon/\slDelta )$
(see section~\ref{gpsi}),
we obtain the following equation for $\ga_{{\rm B}\uparrow}$,
\ber
\ga_{{\rm B}\uparrow}^2e^{2i\vartheta'}+ 
2\ga_{{\rm B}\uparrow}e^{i\vartheta'} \frac{u_{\uparrow}}{t_{\uparrow}t_{\downarrow} } +1=0
\eer
with
$u_{\uparrow}=
\sin \left({\scriptstyle \frac{z}{ \varepsilon_d}} +\vartheta_+ + \slPsi \right)
+ r_{\uparrow}r_{\downarrow}
\sin \left({\scriptstyle \frac{z}{\varepsilon_d}} + \vartheta_- - \slPsi \right)$.
For $\gamma_{{\rm B}\downarrow} $ an analogous equation holds, 
with the quantity
$u_{\downarrow}=
\sin \left({\scriptstyle \frac{z}{\varepsilon_d}} -\vartheta_+ + \slPsi \right)
+ r_{\uparrow}r_{\downarrow}
\sin \left({\scriptstyle \frac{z}{\varepsilon_d}} -\vartheta_- - \slPsi \right)$.
Finally, for the particle-hole conjugated coherence amplitude one obtains
$\gb_{{\rm B}\downarrow} e^{-i\vartheta'}= -\ga_{{\rm B}\uparrow}e^{i\vartheta'} $. 
Thus, the pairing amplitude is given by
\ber
f_{{\rm B}\sigma } = -2\pi i \frac{\ga_{{\rm B},\sigma }}{1-\ga_{{\rm B},\sigma}^2e^{2i\vartheta'}}
= \pi \frac{t_{\uparrow}t_{\downarrow}e^{-i\vartheta'}}{\sqrt{(t_{\uparrow}t_{\downarrow})^2-u_{\sigma}^2}} 
\eer
for $|u_{\sigma}|\le t_{\uparrow}t_{\downarrow}$, and by
\ber
f_{{\rm B}\sigma } 
= i\pi \frac{t_{\uparrow}t_{\downarrow}\sign(u_{\sigma}) }{\sqrt{
u_{\sigma}^2- (t_{\uparrow}t_{\downarrow})^2 }} e^{-i\vartheta'}
\eer
for $|u_{\sigma}|> t_{\uparrow}t_{\downarrow}$.
This characteristic spin dependence of the pairing correlations has been
discussed recently in Ref.~\onlinecite{linder09}, where it was shown that a
change in the symmetry of the pairing correlations near the chemical potential
takes place as function of $\vartheta_{\rm N}$. A more detailed discussion
will be provided in a future publication.\cite{linder09a}

\section{Application II: Superconductor/Half-metal hybrid structure}
\label{sectappl}

\subsection{Interface scattering matrix}

Next, we consider as application an interface between a superconductor
and a completely polarized ferromagnet, a half metal, in the ballistic
limit. Each trajectory
in the superconductor has a spin degeneracy, whereas in the half metal
the spin for each trajectory is fixed.
Following Ref.~\onlinecite{eschrig08},
we write the scattering matrix in singular value decomposition as
\begin{eqnarray}
{\bfm S}= 
\left(
\begin{array}{cc} \hat U_{\rm S} & 0\\ 0& U_{\rm F} \end{array}
\right)
\left( \begin{array}{cc} 
\hat{r}_{{\rm S}} &\ket{t}\\ \bra{t} & -r_{{\rm F}}
\end{array} \right)
\left(
\begin{array}{cc} \hat V^\dagger_{\rm S} & 0\\ 0& V^\dagger_{\rm F} \end{array}
\right)
\nonumber
\end{eqnarray}
where $\ket{t}$ and $\bra{t}$ are the transmission amplitudes, and
$\hat{r}_{\rm S}=\sqrt{1- \ket{t} \bra{t} }$ and $r_{\rm F}=\sqrt{1-
\bra{t}\cdot \ket{t} }$ are the reflection amplitudes.
The phase-matrices on the left and on the right can be written as
$\hat U_{\rm S}=e^{i\left( \psi_u +\frac{\vartheta_u}{2} \vec{m}_u\vec{\sigma }\right)}$,
$\hat V_{\rm S}^{\dagger }=e^{i\left( \psi_v+\frac{\vartheta_v}{2} \vec{m}_v\vec{\sigma } \right)}$,
$U_{\rm F}=e^{i \psi_{\underline{u} }}$, $V_{\rm F}^{\dagger }=e^{i \psi_{\underline v}}$,
and the singular values are determined by the matrices
\begin{eqnarray}
\hat{r}_{\rm S}&=&
\left( \begin{array}{cc} r&0 \\ 0&1 \end{array} \right), 
\qquad \ket{t}=
\left( \begin{array}{c} t \\ 0 \end{array} \right), \\
r_{\rm F}&=&r, \qquad \qquad \; \;
\bra{t}= 
\left( \begin{array}{cc} t & 0 \end{array} \right)
\end{eqnarray}
with $r=\sqrt{1-t^2}$.
The quantization axis is the direction of the magnetization in
the half metal, $\vec{M}$, which we chose as the $z$-axis. 
The directions
$\vec{m}_i$ are determined by the interface properties, and do not 
necessarily coincide with that of the half metal. 

We now make the simplifying model assumption
$\vec{m}_u=\vec{m}_v\equiv \vec{m}$. 
We write $m_x=\sin \alpha \cos \phi$, $m_y=\sin \alpha \sin \phi $,
$m_z=\cos \alpha $, and for the bulk magnetization $M_z=M$, $M_x=M_y=0$.
Because we consider singlet superconductors we have the freedom
to choose a spin quantization axis inside the superconductors
in a convenient way. The most convenient choice is along the
interface magnetic moment $\vec{m}$.
The spin rotation matrix between the quantization axis in the
superconductor and in the half metal is
$\hat U_m=e^{-i \frac{\alpha}{2}\vec{e}_\perp\vec{\sigma}}$ with
$\vec{e}_\perp = (\vec{m} \times \vec{M})/(M \sin \alpha )$. 
In this representation 
$\hat U_m\hat U_{\rm S}\hat U_m^\dagger =e^{i\frac{\vartheta_u}{2}\sigma_z}$
and 
$\hat U_m\hat V_{\rm S}^{\dagger }\hat U_m^\dagger=e^{i\frac{\vartheta_v}{2}\sigma_z}$ become
spin-diagonal. 
Because in quasiclassical approximation only the envelope of the wave
is relevant, we are furthermore allowed to drop all spin-independent 
phases in the scattering matrix (except for a possible phase $\vartheta'$
analogous to that in the last subsection, arising from an internal flux; 
one can prove that all other spin-scalar phases do not enter the final expressions).
This leads to the scattering matrix in the new frame,\cite{eschrig08}
\begin{eqnarray}
&&{\bfm S}
\equiv 
\left( \begin{array}{cc} 
\hat{R}_{\rm S} &\ket{T}\\ \bra{T} & -R_{\rm F}
\end{array} \right) \\
&&= 
\left(
\begin{array}{cc} e^{i\frac{\vartheta_u}{2}\sigma_z} \hat U_m& 0\\ 0& 1 \end{array}
\right)
\left( \begin{array}{cc} 
\hat{r}_{\rm S} &\ket{t}\\ \bra{t} & -r_{\rm F}
\end{array} \right)
\left(
\begin{array}{cc} \hat U_m^\dagger e^{i\frac{\vartheta_v}{2}\sigma_z} & 0\\ 0& 1 \end{array}
\right). \quad
\nonumber 
\end{eqnarray}

\subsection{Josephson geometry}

The Josephson effect in a superconductor/half-metal/superconductor (S/HM/S)
junction has been studied previously both experimentally
\cite{keizer06} 
and theoretically.\cite{eschrig03,volkov03,kopu04,eschrig04,bergeret05,pajovic06,eschrig07,braude07,asano07,linder07,takahashi07,cuoco08,eschrig08,galak08,haltermann08,volkov08,beri09,kalenkov09}
Here we demonstrate how the present formulation of boundary conditions
can be used to simplify analytical expressions within 
the same approximation as in Ref.~\onlinecite{galak08}.
Our formulation is in terms of the microscopic scattering matrix.
Such a scattering matrix cannot in general be obtained by solving Eilenberger's
equations but must be obtained by a full microscopic quantum mechanical treatment of the 
interface.\cite{eschrig07,grein09} This has to be contrasted to the case considered e.g. in Ref.~\onlinecite{volkov08},
where an interface represented by a thin magnetic domain wall is treated with Eilenberger's equations. 
The two approaches are complementary and have non-overlapping ranges of applicability.

\subsubsection{Coherence amplitudes}
\label{SHcoh}

We express the boundary condition it in terms of
the matrices 
\begin{eqnarray}
\mbf{\hat{\gamma }}'
\equiv
\left( \begin{array}{cc} 
\hat{\gamma}'_{\rm S} &\ket{\gamma}'\\ \bra{\gamma}'& \gamma'_{\rm F}
\end{array} \right)=
{\bfm S}
\left( \begin{array}{cc} 
\hat{\gamma}_{\rm S}&0\\0&\gamma_{\rm F}
\end{array} \right)
\tilde{\bfm S}
\end{eqnarray}
with $\hat{\gamma}_{\rm S}$ being a 2$\times $2 spin matrix and $\gamma_{\rm F} $ a scalar, 
and similar notations for the particle-hole conjugated components:
$\hat{\tilde \gamma}_{\rm S}$ and $\tilde \gamma_{\rm F}$.
Explicitely,
\begin{eqnarray}
\hat{\gamma}'_{\rm S} &=& \hat{R}_{\rm S} \hat{\gamma}_{\rm S}  \hat{\tilde R}_{\rm S} +\ket{T}\gamma_{\rm F}\bra{\tilde T} ,\\
\ket{\gamma}' &=& \hat{R}_{\rm S}\hat{\gamma}_{\rm S}\ket{\tilde T} - 
\ket{T} \gamma_{\rm F} \tilde R_{\rm F} ,\\
\bra{\gamma}' &=& \bra{T}\hat{\gamma}_{\rm S}\hat{\tilde R}_{\rm S} - R_{\rm F}\gamma_{\rm F} \bra{\tilde T} ,\\
\gamma'_{\rm F}&=& \bra{T} \hat{\gamma}_{\rm S} \ket{\tilde T} +R_{\rm F}\gamma_{\rm F}\tilde R_{\rm F} .
\end{eqnarray}
Then the boundary conditions, Eqs.~\eqref{GR} and \eqref{GRA}, can be solved
for $\hat \slGamma_{\rm S}$ and $\slGamma_{\rm F}$, leading to
\begin{eqnarray}
\hat \slGamma_{\rm S} &=& \hat \gamma'_{\rm S} + 
\frac{\tilde \gamma_{\rm F} }{ 1- \tilde \gamma_{\rm F} \gamma'_{\rm F} }
\ket{\gamma}' \bra{\gamma}'\\
\slGamma_{\rm F} &=& \gamma'_{\rm F} + \bra{\gamma}'
\left( \hat 1 - \hat{\tilde \gamma}_{\rm S} \hat{\gamma}'_{\rm S} 
\right)^{-1} \hat{\tilde \gamma}_{\rm S}
\ket{\gamma}' .
\label{BCF}
\end{eqnarray}
This gives the outgoing amplitudes in terms of the incoming ones.
The particle-hole conjugated quantities are obtained similarly, with the definition
$\hat{\tilde \gamma}'_{\rm S} = \tilde{\bfm S}^\dagger \hat{\tilde \gamma}_{\rm S} 
{\bfm S}^\dagger $.

We assume singlet superconducting order parameters $\slDelta^{\ret} =|\slDelta|e^{i\chi}i\sigma_y$, allowing us to write for
the bulk coherence functions
$\hat{\gamma}_{\rm S}=\gamma_{\rm S} e^{i\chi }i\sigma_y$ and
$\hat{\tilde{\gamma}}_{\rm S}=\tilde{\gamma}_{\rm S} e^{-i\chi }i\sigma_y$.
It is useful to introduce the parameter
\begin{equation}
P=
\sin \left(\vartheta /2 \right) \sin (\alpha ) / (1+r)
\end{equation}
with the spin mixing angle $\vartheta = \vartheta_u+\vartheta_v$,
that controls the overall magnitude of the proximity effect.
An analytic solution is then given by
\begin{eqnarray}
\label{GammaF}
\slGamma_{\rm F} &=& \frac{ \alpha \gamma_{\rm F} -i\beta \gamma_{\rm S} e^{i(\chi-\phi)} 
}{
\zeta - i\beta \gamma_{\rm F} \tilde \gamma_{\rm S} e^{-i(\chi-\phi) } }
\label{sol}
\end{eqnarray}
where we use the abbreviations
\begin{eqnarray}
\beta&=& Pt^2 (1+r)(1-\gamma_{\rm S}\tilde \gamma_{\rm S})\\
\alpha &= & r^2 + \gamma_{\rm S}^2\tilde\gamma_{\rm S}^2 +2\gamma_{\rm S}\tilde \gamma_{\rm S} r \cos (\vartheta) -\gamma_{\rm S}\tilde\gamma_{\rm S} P^2 t^4 
\\
\zeta&=&1 + \gamma_{\rm S}^2 \tilde \gamma_{\rm S}^2 r^2 +2\gamma_{\rm S} \tilde \gamma_{\rm S}
r\cos (\vartheta) -\gamma_{\rm S} \tilde \gamma_{\rm S} P^2 t^4 \qquad 
\end{eqnarray}
assuming that all incoming coherence amplitudes at the superconducting
side are singlet.
The full solutions of the boundary conditions in the superconductor
can also be obtained analytically and are given in Appendix~\ref{Full}.
Note that the geometric angle $\phi $ that determines the direction
of the interface magnetic moments enters only in combination
with the superconducting order parameter phases. 
Thus, it leads to simple shifts in the current phase relation.\cite{eschrig07,braude07}
In the following, we include $\phi$ into renormalized superconducting 
phases $\chi $ in order to simplify notation, i.e. we define 
$\chi'_1=\chi_1-\phi_1 $, $\chi'_2=\chi_2-\phi_2$
for the two superconducting banks (indices 1 and 2).

\subsubsection{Josephson current}

The equations for the coherence amplitude in a point in the middle of the 
half metal of an S/HM/S junction,
for positive ($\gamma_+$) and negative $(\gamma_-)$ directions,
can be obtained 
by expressing $\slGamma_{F1}$ and $\slGamma_{F2}$  in terms of $\gamma_{F1}$ and
$\gamma_{F2}$ using the boundary conditions Eq.~\ref{GammaF} for each interface,
and solving the transport equations in the half metal with the results
$\gamma_+=y\slGamma_{F1} $, $\gamma_{F2}=y\gamma_+ $,
$\gamma_-=y\slGamma_{F2} $, and $\gamma_{F1}=y\gamma_- $,
where $y= e^{-\frac{\epsilon_n L}{\hbar v_{Fx} } }$,
and $v_{Fx} $ is the component of the Fermi velocity in the half metal perpendicular to
the interfaces.
This leads for a symmetric setup to
\begin{eqnarray}
\gamma_+ &=& y \cdot
\frac{ \alpha  y \gamma_- -i \beta \gamma_{\rm S} e^{i\chi'_1}}{
\zeta-i\beta y\tilde \gamma_{\rm S} e^{-i\chi'_1}\gamma_- }\\
\gamma_- &=& y\cdot
\frac{ \alpha  y \gamma_+ -i \beta \gamma_{\rm S} e^{i\chi'_2}}{
\zeta-i\beta y\tilde \gamma_{\rm S} e^{-i\chi'_2}\gamma_+ } .
\end{eqnarray}

In principle the amplitudes $\gamma_{\rm S}$ and $\tilde\gamma_{\rm S}$ must be
obtained by solving self-consistently for the order parameter
suppression near the interface.
Here, we will however neglect this effect and assume that the bulk solution
\begin{eqnarray}
\gamma_{\rm S}&=& -\tilde \gamma_{\rm S} =
i |\slDelta |/(\epsilon_n + \slOmega_n ), \quad
\slOmega_n=\sqrt{|\slDelta |^2+\epsilon_n^2 } \qquad
\end{eqnarray}
is present all the way to the interface.
This approximation becomes exact in the limit of small $t$ and
$\vartheta $.
Note that in this case 
$1-\gamma_{\rm S}^2= 2\slOmega_n/(|\epsilon_n|+\slOmega_n)$ is even and
$1+\gamma_{\rm S}^2= 2\epsilon_n/(|\epsilon_n|+\slOmega_n)$ is odd in $\epsilon_n$.
One obtains 
\begin{eqnarray}
\frac{g_{+}-g_{-}}{2} &=& -\frac{i \pi}{2} \; \left(\frac{1+\tilde \gamma_+ \gamma_+ }{1- \tilde \gamma_+ \gamma_+} - \frac{1+\tilde \gamma_- \gamma_- }{1- \tilde \gamma_- \gamma_-}\right) \nonumber \\
&=&-i \pi \;\frac{1}{2}\left[ 
\left( 1-\frac{4}{p_{+}^2 }\right)^{-\frac{1}{2}} - \left(1-\frac{4}{p_{-}^2 }\right)^{-\frac{1}{2}} \right]
,\nonumber \\
\end{eqnarray}
by solving the equations $\ga_\pm^2 + p^{\, }_\pm \ga^{\,}_\pm +1=0$.
Here,
\begin{eqnarray}
\label{ppm}
\frac{1}{p_{\pm }}=
\frac{
i\beta \gamma_{\rm S} y \left(\zeta e^{\mp i \frac{\chi }{2}}+\eta y^2 e^{\pm i\frac{\chi}{2}}\right) 
}{
\left( \zeta^2-\eta^2 y^4\right)\pm 2i(\beta\gamma_{\rm S}y)^2 \sin (\chi)
}
\end{eqnarray}
with $\chi=\chi'_2-\chi'_1$.
Note that $\zeta- \eta = t^2(1-\ga_{\rm S}^2 \gb^2_{\rm S})$.

Using this, one can show that for Matsubara frequencies $p_-=p_+^\ast $.
Consequently, the Josephson current is given in terms of these quantities by,
\begin{eqnarray}
\label{Jc}
j= eN_{\rm F}v_{\rm F} \cdot 2\pi T \sum_{\epsilon_n>0}\;\mbox{Im} 
\left\langle \mu
\left(1-\frac{4}{p_{+}(\epsilon_n)^2 }\right)^{-\frac{1}{2}} \right\rangle_{\rm FS_+}
.
\end{eqnarray}
where $\mu =\cos(\theta_p) $, and $\theta_p$ is the impact angle ($\mu=1$ for normal impact).
Here, $v_{\rm F}$ and $N_{\rm F}$ are the Fermi velocity and the density of states at the Fermi level
in the normal state of the half metal, respectively.

We obtain the corresponding Josephson current for the case of the half metal
replaced by a normal metal if we replace $\beta = it^2$, 
$\eta = r^2+\gamma_{\rm S} \tilde \gamma_{\rm S}$,
$\zeta = 1+r^2 \gamma_{\rm S} \tilde \gamma_{\rm S}$ and add a spin degeneracy factor 2.

The normal state boundary resistance of the symmetric S/HM/S Josephson junction 
with area $A$ is given by
\begin{eqnarray}
\label{rn}
\frac{1}{R_{\rm N}A}&=&\frac{j_\perp }{V}=
e^2 N_{\rm F} v_{\rm F} \Big\langle \mu \frac{t^2}{2-t^2}\Big\rangle_{\rm FS+}\\
&=&
e^2 \int\limits_{(\vec{v}_{\rm F} \vec{e}_\perp)>0} \frac{(dp_{\rm F})}{|\vec{v}_{\rm F}|} (\vec{v}_{\rm F} \vec{e}_\perp) \;
\frac{t(\vec{p}_{||})^2}{2-t(\vec{p}_{||})^2} \qquad\\
&=& e^2 \int (dp_{||}) \frac{t(\vec{p}_{||})^2}{2-t(\vec{p}_{||})^2}
\label{rna_cc}
\end{eqnarray}
with $(dp_{\rm F})=d^{D-1}p_{\rm F}/(2\pi \hbar)^D$ for $D$ dimensions, and
$\vec{p}_{||}=\vec{p}_{\rm F} \cdot \vec{e}_{||}$.\cite{footRn}
For an S/N/S junction an additional factor 2 has to be added on the right hand sides.

\begin{figure}[t]
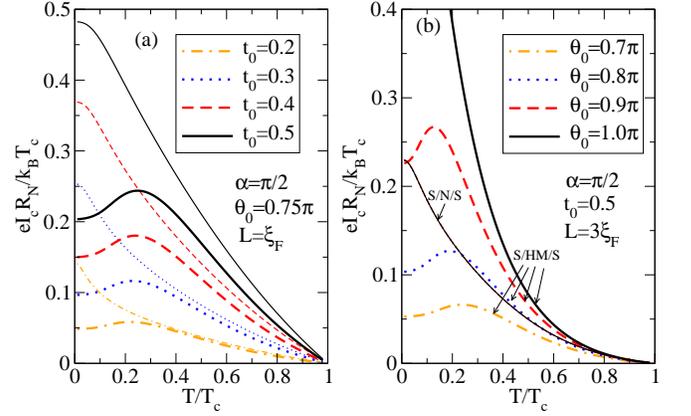

\includegraphics[width = 0.49\columnwidth]{Jc_T1.eps}
\includegraphics[width = 0.49\columnwidth]{Jc_T1_theta.eps}
\caption{\label{Jc_T}
(Color online)
Critical Josephson current $I_{\rm c}$ multiplied with the normal state resistance
$R_{\rm N}$ for an S/HM/S Josephson junction with
magnetic interfaces (thick lines) and for an
S/N/S junction with non-magnetic interfaces (thin lines). 
In (a) for both cases, the transmission amplitudes 
$t_0$ are varied from 0.2 to 0.5.
The spin-mixing angle for normal impact is $\vartheta_0=0.75 \pi$, 
and the junction length $L$ is equal
to the coherence length in the half metal, $\xi_{\rm F}=v_{\rm F}/2\pi T_{\rm c}$.
In (b) we fix $L=3\xi_{\rm F}$, $t_0=0.5$,
and vary for the S/HM/S junction the spin mixing angle.
For (a) and (b) the interface spin misalignment angle is $\alpha=\pi/2$. 
}
\end{figure}
In Fig.~\ref{Jc_T}(a) we show results obtained with Eq.~\ref{Jc}. Shown
is the critical Josephson current multiplied with the normal state resistance
obtained by Eq.~\ref{rn}. For definiteness we present results for
identical isotropic Fermi surfaces on both sides of the interface and 
for the dependence of the transmission amplitude $t$ on the
impact angle $\theta_p$ (measured from the surface normal) 
appropriate for a $\delta $-potential,
$t(\theta_p)=t_0 \cos \theta_p/\sqrt{1-t_0^2 \sin^2 \theta_p}$,
where $t_0$ is the transmission for normal impact.
For the spin mixing angle $\vartheta $ we assume a dependence
$\vartheta= \vartheta_0 \cos(\theta_p)$. 
For comparison we also show the corresponding values for a superconductor/normal-metal/superconductor (S/N/S) Josephson junction.
As can be seen, the supercurrent through a half metal can be of a similar
magnitude as through a normal metal, provided the parameter $P$ is of order one.

In fact, as can be seen in Fig.~\ref{Jc_T}(b), the $I_{\rm c}R_{\rm N}$-product
can exceed that for an analogous S/N/S junction. 
The reason for this enhancement are current carrying Andreev bound states
below the gap energy, that are discussed further below.
We show for several values of $\vartheta_0$ the $I_{\rm c}R_{\rm N}$-product in comparison
with that for a non-magnetic S/N/S Josephson junction 
with the same transmission probability and same length. With increasing $\vartheta_0$
the magnitude of the effect increases, and the maximum in the temperature dependence
moves to lower temperatures. 

In fact, for the special case that
$P=1$ (i.e. $t=1$, $\vartheta=\pi$, $\alpha=\pi/2$)
for all Fermi surface points, the maximum becomes unobservable because it
moves to zero temperature, as has been noted also in Ref.~\onlinecite{galak08}.
In this case, furthermore, we have $\beta=\zeta=(1-\ga_{\rm S}\gb_{\rm S})$, 
$\eta=-\ga_{\rm S}\gb_{\rm S} (1-\ga_{\rm S}\gb_{\rm S})$ for the S/HM/S junction, and $\beta=i$, $\zeta=1$, $\eta= \ga_{\rm S}\gb_{\rm S}$ for the S/N/S junction.
Consequently, after canceling the common factor $(1-\ga_{\rm S}\gb_{\rm S})$
in Eq.~\ref{ppm} for the S/HM/S junction, it is seen that $1/p_\pm$ at
phase $\chi$ for the S/HM/S junction coincides with $1/p_\pm $ at phase $(\chi+\pi)$
for the S/N/S junction. This proves that the $I_{\rm c} R_{\rm N}$-product
for $P=1$ is equal to that for the corresponding S/N/S junction, and the 
corresponding current phase relations are shifted by $\pi$.
This result is in agreement with the findings
in section III.D of Ref.~\onlinecite{galak08} for the short and long junction limits,
that were obtained within the more general Gor'kov formalism.

We caution however, that the suppression of the singlet order parameter at the
interface cannot be neglected for $P$ close to $1$,
unless a strong Fermi surface mismatch is present
(in which case the transmission is reduced due to the Fermi velocity mismatch),
and self-consistent calculations must be performed as done in
Ref.~\onlinecite{eschrig03}.

\subsubsection{Local density of states}

We now proceed to calculate the local density of states as function of
energy. For this we need to perform an analytical continuation to the
real energy axis.  We define in this case $y=e^{izL/\mu v_{\rm F}}$ and 
$\gamma_{\rm S}=-\tilde \gamma_{\rm S} = -|\slDelta |/(z+i\sqrt{|\slDelta|^2-z^2})$ with
$z=\epsilon+i0^+ $.
The momentum resolved density of states is then given in the center of the
half metal by,
\begin{eqnarray}
\frac{N_\pm}{N_{\rm F}}&=&-\; \frac{1}{\pi }\mbox{Im} g^\ret_{\pm} = \mbox{Re} \frac{1+\tilde \gamma_\pm\gamma_\pm}{1-\tilde \gamma_\pm\gamma_\pm} 
\nonumber \\
&=&\mbox{Re} \left[ \left( 1-\frac{4}{p_{\pm}(\epsilon )^2 }  \right)^{-\frac{1}{2}}\right]. 
\end{eqnarray}
The local density of states is obtained as 
\beq
N(\epsilon )= \langle N_+ +N_- \rangle_{\rm FS_+}.
\eeq
\begin{figure}[b]
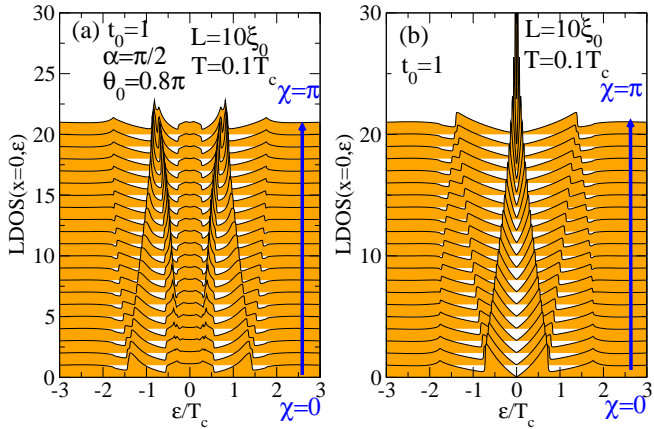

\includegraphics[width = 0.49\columnwidth]{DOS13_2.eps}
\includegraphics[width = 0.49\columnwidth]{DOS13a_2.eps}
\caption{\label{DOS13}
(Color online)
Local density of states in the center of a current biased high-transmissive
symmetric Josephson junction
for (a) an S/HM/S junction  and (b) an S/N/S junction. In both cases the
phase difference over the junction is varied from 0 to $\pi$. The remaining
parameters are indicated. 
}
\end{figure}
In Fig.~\ref{DOS13} we compare the local density of states (LDOS) for an S/HM/S-junction
and an S/N/S-junction in the high-transmission limit for a symmetric setup.
For clarity of presentation we have vertically shifted the curves with respect to each other.
The junctions are current biased, and the phase difference varies in both
cases from 0 to $\pi$ as indicated. For the S/N/S-junction the well-known
Andreev-Saint-James states\cite{andreev64,saint64}
are seen (for a review see Ref.~\onlinecite{deutscher05}) with a reduction of the  LDOS
at low bias except for the case of $\chi=\pi$, when a zero bias bound state
is present. In contrast, for the S/HM/S-junction there is a
low-energy band of bound states.
This behavior has already been noted in Ref.~\onlinecite{eschrig03}.
Note that $\chi=\pi $ is the equilibrium phase of the S/HM/S junction.\cite{eschrig03}
The dispersion of the Andreev peaks in the spectra with $\chi $ indicates the
direction of the current that is carried by them. For the S/N/S-junction
the lowest bias peak dominates, that carries current in positive direction,
whereas for the S/HM/S-junction the low-energy band is responsible for
the low-temperature anomaly $J_{\rm c}(T)$, and the next higher band carries
most of the current, that is in
negative direction, in accordance with the $\pi $-junction behavior.

The half-width $W_{1/2}$ of the low-energy band varies 
with the interface parameters, with the impact angle, with the phase
difference $\chi $, with temperature, and with junction length. 
In general the width of the low-energy band is larger for $\chi=0$ than for $\chi=\pi$. 
\begin{figure}[b]
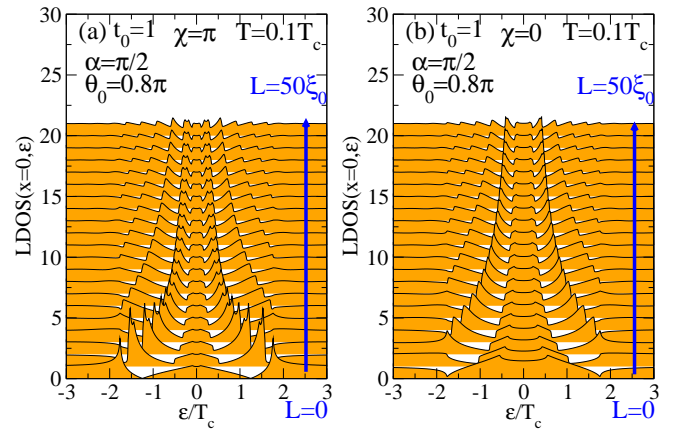

\includegraphics[width = 0.49\columnwidth]{DOS12_1.eps}
\includegraphics[width = 0.49\columnwidth]{DOS12a_1.eps}
\caption{\label{DOS12}
(Color online)
Local density of states in the center of a current biased high-transmissive
symmetric S/HM/S Josephson junction, for (a) a
phase difference over the junction of $\pi$ and (b) of $0$. 
In both cases the junction length is varied from the short junction limit
to $L=50\xi_{\rm F}$.
The remaining parameters are indicated.
}
\end{figure}
In Fig.~\ref{DOS12}, we show its dependence on the junction length
for (a) a $\pi$-junction and (b) a zero-junction.
In the short-junction limit the half-width for $t=1$ 
is given by $W_{1/2}(\chi=\pi)=|\slDelta | (\sqrt{2-P^2} -P)/2$ and
$W_{1/2}(\chi=0)=|\slDelta | \sqrt{1-P^2}$. In the limit of small $P$ (but still
$t=1$)
we obtain $W_{1/2}(\chi=\pi ) \to |\slDelta |/\sqrt{2} $ and $W_{1/2}(\chi=0)\to |\slDelta |$.
For the special case $P=1$ the spectra are equal to those for an S/N/S-junction
with the junction phase shifted by $\pi$ (see the corresponding discussion in the last subsection), i.e. the low energy band vanishes in the limit $P\to 1$
for a $\pi $ junction and
a zero energy bound state appears for a zero-junction.
In general, as $P$ varies with $\vartheta $ and thus with the impact angle, the
width of the low-energy band in Figs.~\ref{DOS13}(a) and \ref{DOS12} is a superposition
for many different $P$. The overall width of the low-energy band
is set by the values for smallest $P$, the kink-features closer to the
chemical potential correspond to the largest $P$ for
trajectories with normal impact.

\begin{figure}[t]
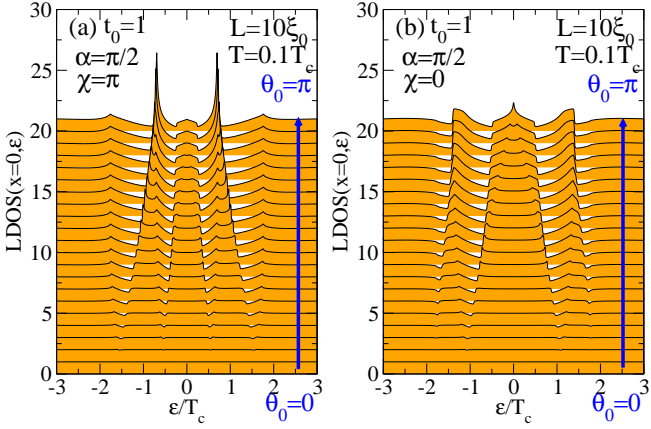

\includegraphics[width = 0.49\columnwidth]{DOS11_1.eps}
\includegraphics[width = 0.49\columnwidth]{DOS11a_1.eps}
\caption{\label{DOS11}
(Color online)
Local density of states in the center of a current biased high-transmissive
symmetric S/HM/S Josephson junction, for (a) a
phase difference over the junction of $\pi$ and (b) of $0$. 
In both cases the spin mixing angle $\vartheta_0$
is varied from 0 to $\pi$. The remaining
parameters are indicated.
}
\end{figure}
In Fig.~\ref{DOS11} we show results for the variation of the LDOS with with the
spin-mixing angle $\vartheta_0$ for (a) a $\pi$-junction and (b) a
zero-junction. For the zero-junction a peak appears at high $\vartheta_0$, that
is a signature of the zero bias bound state for normal impact when $P=1$.
For smaller spin-mixing angles in general the structures
get smeared out, and a set of energy bands separated by narrower suppressions
of the LDOS remains.

\subsection{Point contact geometry}

\subsubsection{Distribution functions}

For the distribution functions, we introduce the notation,
\begin{eqnarray}
\hat{\bfm x }'=
\left( \begin{array}{cc} 
\hat{x}'_{\rm S} &\ket{x}'\\ \bra{x}'& x'_{\rm F}
\end{array} \right)
=
{\bfm S}^\ret
\left( \begin{array}{cc} 
\hat{x}_{\rm S}&0\\0&x_{\rm F}
\end{array} \right)
{\bfm S}^\adv
\end{eqnarray}
with ${\bfm S}^\adv =({\bfm S}^\ret )^\dagger $,
which explicitely gives
\begin{eqnarray}
\hat{x}'_{\rm S} &=& \hat{R}_{\rm S}^\ret \; \hat{x}_{\rm S}   \hat{R}_{\rm S}^\adv +
x_{\rm F} \ket{T}^\ret \bra{T}^\adv \\
\ket{x}' &=& \hat{R}_{\rm S}^\ret \; \hat{x}_{\rm S} \ket{ T}^\adv - 
x_{\rm F} \; R_{\rm F}^\adv \; \ket{T}^\ret \\
\bra{x}' &=& \bra{T}^\ret  \hat{x}_{\rm S} \; \hat{ R}_{\rm S}^\adv  - 
x_{\rm F}\; R_{\rm F}^\ret  \; \bra{ T}^\adv  \\
x'_{\rm F}&=& \bra{T}^\ret   \hat{x}_{\rm S}  \ket{ T}^\adv  +
x_{\rm F}\; R_{\rm F}^\ret \; R_{\rm F}^\adv  .
\end{eqnarray}
with $\hat{R}_{\rm S}^\adv = (\hat{R}_{\rm S}^\ret )^\dagger $, $\ket{ T}^\adv = (\bra{ T}^\ret )^\dagger $, $\bra{ T}^\adv = (\ket{T}^\ret )^\dagger $, $R_{\rm F}^\adv =(R_{\rm F}^\ret )^\ast$.
Then, with the abbreviations
\begin{eqnarray}
\ket{\slGamma }^\ret &=& \ket{\gamma }' (1- \tilde \gamma_{\rm F} \gamma'_{\rm F} )^{-1} \\
\bra{\slGamma }^\ret &=& \bra{\gamma }' 
\left( \hat 1 - \hat{\tilde \gamma}_{\rm S} \hat{\gamma}'_{\rm S} \right)^{-1},
\end{eqnarray}
and $\ket{\slGamma }^\adv = (\bra{\sltilde \slGamma }^\ret )^\dagger $,
$\bra{\slGamma }^\adv = (\ket{\sltilde \slGamma }^\ret )^\dagger $,
the explicit boundary conditions, Eq.~\eqref{X}, for the distribution functions read,
\begin{eqnarray}
\hat X_{\rm S} &=& \hat x'_{\rm S} + 
\gamma_{\rm F}^\adv
\ket{x}' 
\bra{\sltilde \slGamma }^\adv 
+ 
\tilde \gamma_{\rm F}^\ret  
\ket{\slGamma}^\ret 
\bra{x}'
\nonumber \\
&&\qquad +
(\tilde \gamma^\ret_{\rm F} x'_{\rm F} \gamma_{\rm F}^\adv -\tilde x_{\rm F}) 
\ket{\slGamma}^\ret 
\bra{\sltilde \slGamma }^\adv \\
\label{pcX}
X_{\rm F} &=& x'_{\rm F} + 
\bra{x}' \; \hat \gamma_{\rm S}^\adv \; \ket{\sltilde \slGamma }^\adv +
\bra{\slGamma}^\ret \; \hat {\tilde \gamma}_{\rm S}^\ret \; \ket{x}'
\nonumber \\
&&\qquad +\bra{\slGamma}^\ret \; (\hat {\tilde \gamma}^\ret_{\rm S} \hat x'_{\rm S} \hat \gamma_{\rm S}^\adv -\hat{\tilde x}_{\rm S}) \; \ket{\sltilde \slGamma }^\adv 
\label{BCKel}
\end{eqnarray}
Here, $\hat \gamma_{\rm S}^\adv = (\hat {\tilde \gamma}_{\rm S}^\ret )^\dagger $, $\gamma_{\rm F}^\adv =(\tilde \gamma_{\rm F}^\ret )^\ast $.

\subsubsection{Point contact spectra }

Superconductor/half-metal point contact spectra have been studied
experimentally in a number of cases.\cite{soulen98,desisto00,ji01,angu01,parker02,woods04,dyachenko06,yates07,krivoruchko08,bocklage07}
However, the analysis in all these studies did not include the effect of
spin active interface scattering.
Here, we show how such effects can be taken into account in a
ballistic point contact.
We assume incoming solutions to be in equilibrium. The treatment in terms of
coherence and distribution functions can be simplified considerably by
using the symmetries described in Appendix~\ref{gauge2}.
Proceeding along the lines described there,
we introduce anomalous distribution functions by
$\hat g^K[x,\tilde x]-\hat g^K[x_0,\tilde x_0]=
\hat g^\ret \hat F_0 - \hat F_0 \hat g^\adv $, with 
\begin{eqnarray}
\hat F_0= \left( \begin{array}{cc} F_0 &0 \\ 0 &-\tilde F_0 \end{array} \right)
\end{eqnarray}
and $x_0=x -(F_0 + \gamma^\ret \tilde F_0 \tilde \gamma^\adv )$.
We use for $F_0$ the equilibrium distribution function in the superconductor.
Then, 
the incoming anomalous distribution functions $x_{S,0} $ and $\tilde x_{S,0}$ 
in the superconductor are zero. 
For the half metal we have 
$x_{\rm F}=F+\gamma^\ret_{\rm F} \tilde F \tilde \gamma^\adv_{\rm F}$ with
$F=\tanh [(\epsilon-eV)/2T]$ and
$\tilde F= -\tanh[(\epsilon+eV)/2T]$, and consequently
$x_{F,0}=(F-F_0)+\gamma^\ret_{\rm F} (\tilde F -\tilde F_0) \tilde \gamma^\adv_{\rm F}$.
Furthermore, for a ballistic point contact the incoming coherence functions
on the half-metallic side are zero, $\gamma_{\rm F}=\tilde \gamma_{\rm F}=0$.
From here on we drop the index ``0'' for all distribution
functions in order to not overload the notation, and keep in mind that
they are all anomalous.

Substituting all this into Eq.~\eqref{pcX}, one arrives at
\begin{eqnarray}
&&X_{\rm F}-x_{\rm F} = 
-\frac{x_{\rm F}t^2}{
|\zeta^\ret |^2 }\Big\{
(1-\slPi_2) \left[ 1+r^2\slPi_2 -r\cos (\vartheta ) \slSigma_2
\right. \Big. \nonumber \\ 
&&\qquad \quad +
\Big. \left.
r\cos (\alpha) \sin (\vartheta) \slDelta_2 \right]
\Big.  \nonumber \\ 
&&\qquad \quad +
\Big. 
t^2P^2 (1+r)\left[ \slSigma_2 (1+r\slPi_2) +2 (1+r)\slPi_2 \right]
\Big\} \nonumber \\
&&-\slGamma_{\rm F}^\ret \tilde x_{\rm F} \sltilde \slGamma_{\rm F}^\adv =
-\frac{\tilde x_{\rm F}t^4}{
|\zeta^\ret |^2} P^2 (1+r)^2|\gamma_{\rm S}^\ret |^2
\left[1+\slSigma_2 + \slPi_2 \right] \nonumber \\
\end{eqnarray}
with 
\begin{eqnarray}
\slPi_2&=& (\gamma_{\rm S}^\ret \tilde\gamma_{\rm S}^\ret )(\gamma_{\rm S}^\adv \tilde \gamma_{\rm S}^\adv )^2
= |\gamma_{\rm S}^\ret  \tilde \gamma_{\rm S}^\ret |^2 \\
\slSigma_2&=& -(\gamma_{\rm S}^\ret \tilde \gamma_{\rm S}^\ret +\gamma_{\rm S}^\adv \tilde \gamma_{\rm S}^\adv )
=-2\mbox{Re} (\gamma_{\rm S}^\ret \tilde \gamma_{\rm S}^\ret ) \\
\slDelta_2&=& -\frac{1}{i}(\gamma_{\rm S}^\ret \tilde \gamma_{\rm S}^\ret -\gamma_{\rm S}^\adv \tilde \gamma_{\rm S}^\adv )
=-2\mbox{Im} (\gamma_{\rm S}^\ret \tilde \gamma_{\rm S}^\ret ),
\end{eqnarray}
and we have used the notation
$\hat \gamma_{\rm S}^\ret = \gamma_{\rm S}^\ret  i\sigma_y$ and 
$\hat \gamma_{\rm S}^\adv = \gamma_{\rm S}^\adv  i\sigma_y$, meaning that 
$\tilde \gamma_{\rm S}^\adv =-(\gamma_{\rm S}^\ret )^\ast $.
These expressions are still general and we only made use of zero incoming
$\gamma_{\rm F}$, $\tilde \gamma_{\rm F}$, $x_{S,0}$, and $\tilde x_{S,0}$.

The current density from the half metal to the superconductor is given in terms
of the anomalous distribution functions by,
\begin{eqnarray}
j=eN_{\rm F}v_{\rm F} \int \frac{d\epsilon}{2}  \; \Big\langle \mu (X_{\rm F}-x_{\rm F}-\slGamma_{\rm F}^\ret \tilde x_{\rm F} \sltilde \slGamma_{\rm F}^\adv )\Big\rangle_{\rm FS_+}
\end{eqnarray}
with $x_{\rm F}= F-F_0=[\tanh \frac{\epsilon-eV}{2T}-\tanh \frac{\epsilon}{2T}]$, and
$\tilde x_{\rm F}=\tilde F-\tilde F_0= -[\tanh\frac{\epsilon+eV}{2T}-\tanh \frac{\epsilon}{2T}]$.
The current density can be written as
\begin{eqnarray}
j&=&eN_{\rm F}v_{\rm F} \int_{-\infty }^{\infty } \frac{d\epsilon}{4}  \; j_\epsilon
\left(\tanh \frac{\epsilon+eV}{2T}-\tanh \frac{\epsilon-eV}{2T}\right)
\nonumber \\
\end{eqnarray}
with spectral current kernels $j_\epsilon $.
The normal state boundary resistance is $1/R_{\rm N}A=j_{\rm N}/V=e^2N_{\rm F}v_{\rm F} \langle \mu t^2 \rangle_{\rm FS_+}$.

\begin{figure}[t]
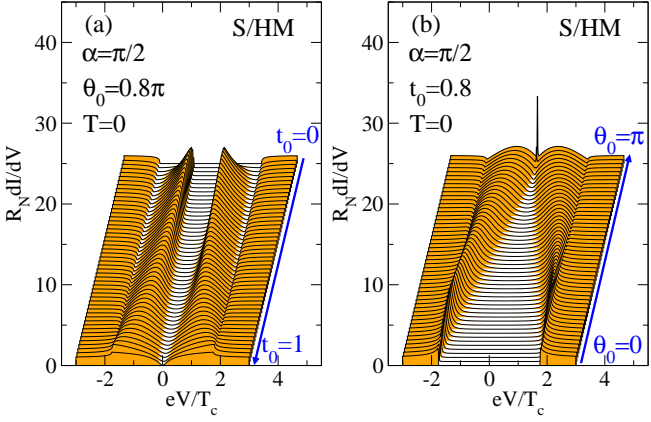

\includegraphics[width = 0.49\columnwidth]{ADOS12.eps}
\includegraphics[width = 0.49\columnwidth]{ADOS13.eps}
\caption{\label{ADOS12}
(Color online)
Point contact spectra for an S/HM contact.
In (a) the transmission $t_0$ varied from 0 to 1, and in (b)
the spin-mixing angle $\vartheta_0$ is varied from 0 to $\pi $. Both quantities
depend on the quasiparticle impact angle as discussed in the text.
The remaining parameters are indicated. 
}
\end{figure}
Further simplifications arise for incoming homogeneous distribution functions,
when $\tilde \gamma_{\rm S}^\ret =-\gamma_{\rm S}^\ret $. Then,
noting that for $|\epsilon |< |\slDelta |$ we have $\slPi_2=1$, it follows that
\begin{eqnarray}
X_{\rm F}-x_{\rm F} &=& 
-\frac{x_{\rm F}t^4}{|\zeta^\ret |^2} P^2 (1+r)^2(\slSigma_2 +2) 
\\
-\Gamma_{\rm F}^\ret  \tilde x_{\rm F} \tilde \Gamma_{\rm F}^\adv  &=&
-\frac{\tilde x_{\rm F}t^4}{|\zeta^\ret |^2} P^2(1+r)^2 (\slSigma_2 +2)
\end{eqnarray}
leading to the Andreev spectral current
\begin{equation}
j_\epsilon =\left\langle \frac{\mu \cdot
4P^2 t^4(1+r)^2 \cdot
\big( 1+\mbox{Re} [(\gamma_{\rm S}^\ret )^2] \big)
}{\Big|1+(\gamma_{\rm S}^\ret )^4 r^2 -2(\gamma_{\rm S}^\ret )^2 r \cos(\vartheta ) +(\gamma_{\rm S}^\ret )^2 P^2 t^4\Big|^2} 
\right\rangle_{\!\!\!\!\rm FS_+}\!\!\!\!\!\!.
\end{equation}
For zero misalignment of the interface moments,
$\alpha=0$ leads to $P=0$, and there is no Andreev current.
For $|\epsilon | > |\slDelta |$, additional
terms become important, associated with $\slPi_2\ne 1$.
Here, because $\gamma_{\rm S}^\ret $ is real, and thus
$\tilde x_{\rm F}(\epsilon) = -x_{\rm F}(-\epsilon) $,
we obtain 
\begin{eqnarray}
j_\epsilon =
\left\langle 
\frac{\mu t^2 \cdot [1-(\gamma_{\rm S}^\ret )^4 ]
}{1+(\gamma_{\rm S}^\ret )^4 r^2 -2(\gamma_{\rm S}^\ret )^2 r \cos(\vartheta ) +(\gamma_{\rm S}^\ret )^2 P^2 
t^4} 
\right\rangle_{\rm FS_+} &&
\nonumber \\  +
\left\langle 
\frac{\mu \cdot 2P^2t^4 (1+r)^2 \cdot [1+(\gamma_{\rm S}^\ret )^2 ]^2 (\gamma_{\rm S}^\ret )^2 
}{\Big|1+(\gamma_{\rm S}^\ret )^4 r^2 -2(\gamma_{\rm S}^\ret )^2 r \cos(\vartheta ) +(\gamma_{\rm S}^\ret )^2 P^2 t^4\Big|^2} 
\right\rangle_{\!\!\!\!\rm FS_+}  \!\!\!\!\!\!.&&\nonumber \\
\end{eqnarray}

For comparison, we also present the expressions for the
corresponding ${\rm d}I/{\rm d}V$-spectra for a normal metal,
$j_\epsilon =1+\left\langle \mu [
t^4|\gamma_{\rm S}^\ret |^2 -r^2 |1-(\gamma_{\rm S}^\ret )^2|^2]/
|1 -r^2(\gamma_{\rm S}^\ret )^2 |^2 \right\rangle_{\rm FS_+}$.
This gives 
$j_\epsilon =\left\langle 2\mu 
t^4/ |1 -r^2(\gamma_{\rm S}^\ret )^2 |^2 \right\rangle_{\rm FS_+}$,
for $|\epsilon|<|\slDelta |$,
and 
$j_\epsilon =\left\langle \mu 
t^2 [1+(\gamma_{\rm S}^\ret )^2]/ [1 -r^2(\gamma_{\rm S}^\ret )^2]\right\rangle_{\rm FS_+}$
for $|\epsilon|>|\slDelta |$.

In Fig.~\ref{ADOS12} we show representative results for zero temperature S/HM point-contact spectra 
for various transmissions
and spin-mixing angles. In general, there are sub-gap states present in
the spectra except for very small $\vartheta_0$, $\alpha$, or $t_0$.
For the special case $\vartheta_0=\pi $ there is a sharp zero bias
state observable in the spectra. Otherwise, if $\vartheta_0 $ is not close
to $\pi$, ${\rm d}I/{\rm d}V$ is for $T=0$ zero at zero bias and increases quadratic with
the voltage. 
The details of the spectra will depend on the Fermi surface mismatch and
the interface characteristics, in particular the dependence of the various
parameters on impact angle.
We leave a detailed discussion of these issues for a future publication.

\section{Acknowledgments}
I would like to thank for
valuable discussions with 
Jaime Ferrer,
Mikael Fogelstr\"om, 
Tomas L\"ofwander, 
Jim Sauls, 
Anton Vorontsov,
Sungkit Yip, and
Erhai Zhao
about quasiclassical boundary conditions,
and with
Roland Grein,
Juha Kopu,
Tomas L\"ofwander,
Georgo Metalidis, and
Gerd Sch\"on 
about superconductor/half-metal heterostructures.
I also acknowledge 
the hospitality of the Aspen Center for Physics, where
part of this work was done.

\appendix 
\section{Time convolution product}
\label{Notation}

We use extensively the non-commutative $\qt $-product between two functions,
which allows us to formulate the equations independently from the 
representation of the dynamical coordinates (time, energy, mixed). 
In the time domain,
the non-commutative $\qt $-product between two functions
$\op{A}(t,t')$ and $\op{B}(t,t')$ is defined by
\beq
\op{A} \qt \op{B}(t,t') = \int {\rm d}t'' \; \op{A}(t,t'') \op{B}(t'',t'),
\eeq
with the unit element $\hat{\it 1}=\delta(t-t')\hat 1$.
In an energy representation (after a Fourier transform $t\to \epsilon $,
$t'\to \epsilon'$), the product reads
\beq
\op{A} \qt \op{B}(\epsilon,\epsilon') = \int \frac{{\rm d}\epsilon''}{2\pi} \; \op{A}(\epsilon,\epsilon'') \op{B}(\epsilon'',\epsilon'),
\eeq
with the unit element $\hat{\it 1}=\delta(\epsilon-\epsilon')\hat 1$.
In a mixed representation, when performing a Fourier transform
$(t-t') \to \epsilon $, and keeping the time variable $(t+t')/2 \to t$,
the product can be written as
\beq
\label{timeconv}
\op{A} \qt \op{B}(\ep,t) =
\mbox{e}^{\frac{i\hbar }{2}(\partial_\ep^\adv  \partial_t^B-\partial_t^\adv  \partial_\ep^B)}
\op{A}(\ep,t) \op{B}(\ep,t) \quad ,
\eeq
and the unit element is $\hat{\it 1}=\hat 1$.
If one of the factors is both independent of $\ep $ and $t$, the
$\qt $-product reduces to the usual matrix product.
Note that in a mixed representation 
\beq
\ep \qt a - a \qt \ep = i\hbar \partial_t a, \qquad
\ep \qt a + a \qt \ep = 2\ep a.
\eeq
Sometimes (for example when performing a perturbation theory out of the 
equilibrium)
a modified energy representation is useful, where
one performs Fourier transforms 
$(t-t') \to \epsilon $, $(t+t')/2 \to \om $. 
In this case the product reads
\ber 
\op{A} \qt \op{B}(\ep,\om ) &=&
\int\limits_{-\infty}^{\infty} \frac{\mbox{d}\om'}{2\pi }
\frac{\mbox{d}\om''}{2\pi } \delta (\om' + \om'' - \om ) \times \nn \\
&&\times \op{A}(\epp{\ep}{\hbar \om'},\om'' ) \op{B}(\epm{\ep}{\hbar \om''},\om') \quad ,
\eer
and the unit element is $\hat{\it 1}=\delta(\om )\hat 1$.
If $\op{A}(\ep,t)=\op{A}(\ep)$ is independent of $t$ (if
$\op{A}$ is an equilibrium quantity) then
\beq
\op{A} \qt \op{B}(\ep,\om) =
\op{A}(\epp{\ep}{\hbar \om}) \op{B}(\ep,\om) \quad,
\eeq
and, analogously, if $\op{B}$ is an equilibrium quantity
\beq
\op{A} \qt \op{B}(\ep,\om) =
\op{A}(\ep,\om) \op{B}(\epm{\ep}{\hbar \om})\quad. 
\eeq
We also generalize throughout the paper the commutator
\beq
[\op{A},\op{B}]_{\qt } = \op{A}\qt \op{B} - \op{B}\qt\op{A} \; .
\eeq
A useful identity is
\beq
\label{id1}
\left( 1+ a \qt b \right)^{-1} \qt a = 
a \qt \left( 1+ b \qt a \right)^{-1} . 
\eeq

\section{Projectors}\label{appB}

We adopt here the notation of Ref.~\onlinecite{eschrig00}, Appendix B.
Following Shelankov,\cite{shelankov85}
we introduce the following projectors
\ber
\label{proj}
\PPpm = \half \left( \ce \pm \frac{\; 1}{-i\pi} \; \cg \right) 
\eer
with the properties
$\PPp \qt \PPp = \PPp $ , $\PPm \qt \PPm = \PPm $,
$\PPp + \PPm = \ce $, and
$\PPp \qt \PPm = \PPm \qt \PPp = \cz $.
The quasiclassical Green's function is expressed
in terms of $\PPp $ or $\PPm $ by,
\ber
\label{gqra1}
\cg &=& -i\pi \left( \PPp - \PPm \right) 
\, .
\eer
From the normalization condition,
the Keldysh component of the Green's function, $\gqk $, fulfills the
relations
$\Par \qt \gqk \qt \Paa = \op{0}$ and
$\Pbr \qt \gqk \qt \Pba = \op{0}$, which allows a parameterization
by
\beq
\label{gqk2}
\gqk = -2\pi \; i \; \left[ \Par \qt \XX \qt \Pba + \Pbr \qt \YY \qt \Paa \right]\; ,
\eeq
were $\XX$ and $\YY$ are related by symmetry relations. The function
$\XX $ can be chosen in a convenient way.

Analogously, for the linear response to an external perturbation,
the normalization condition leads to
$\Para \qt
\dgqra \qt \Para=\op{0}$ 
and $\Pbra \qt \dgqra
\qt \Pbra=\op{0}$;
as a consequence
the spectral response, $\dgqra $,
can be written as 
\ber
\label{dXXYYra}
\dgqra = 
\mp 2\pi i \left[  \Pa \qt \dWW
\qt \Pb- \Pb \qt \dZZ
\qt \Pa \right]^{\ra} 
\eer
with a suitable parameterization of the functions $\dWW$ and $\dZZ$.

\section{Parameter representations of Projectors }
\label{appC}

The projectors $\Par $ and $\Pbr $ may be parameterized 
by complex spin matrices $\gar $ and $\gbr $
as defined in Appendix C of Ref.~\onlinecite{eschrig00}. Alternatively, we give
here the parameterization in terms of ${\cal G}$, ${\cal F}$, 
$\tilde{\cal G}$, and $\tilde{\cal F}$. We obtain
\ber
\label{par}
\Par = 
\mat \plus {\cal G} & {\cal F}\\ -\tilde {\cal F} &
(1-\tilde {\cal G})
\matend^{\ret} ,
\label{pbr}
\Pbr = 
\mat (1-{\cal G}) & -{\cal F} \\ \tilde{\cal F} & \tilde {\cal G} \matend^{\ret}
.
\eer
\ber
\label{paa}
\Paa = 
\mat (1-\tilde{\cal G}) &-{\cal F}  \\ \tilde{\cal F} & \tilde{\cal G} \matend^{\adv} 
,
\label{pba}
\Pba = 
\mat \plus {\cal G} & {\cal F} \\ -\tilde{\cal F} & (1-\tilde{\cal G})  \matend^{\adv}
.
\eer

\section{Parameter representations of distribution functions}
\label{distpar}
In general the functions $\XX $ and $\YY$ in Eq.~\eqref{gqk2} can be written as
\beq
\label{kelgf1}
\XX = \mat \xa_{11} & \xa_{12} \\ \yb_{12} & \yb_{11} \matend , \quad 
\YY = \mat \ya_{11} & \ya_{12} \\ \xb_{12} & \xb_{11} \matend,
\eeq
taking into account the fundamental symmetry relations for the Keldysh Green's function
through the ``tilde'' particle-hole symmetry relation.
Any choice of the four functions $\xa_{11}$, $\xa_{12}$, $\ya_{11}$, and $\ya_{12}$,
will lead to a valid parameterization of the Keldysh Green's function.
As they parameterize only one free function in $\gqk $ (due to
symmetry relations and normalization condition),
three of the four parameters can be chosen conveniently.
It is customary to require $\xa_{12} = \ya_{12}=0$, leading to the parameterization
\beq
\XX = \mat \xa & 0 \\ 0 & \yb \matend , \quad 
\YY = \mat \ya & 0 \\ 0 & \xb \matend.
\eeq
Three definitions for distribution functions have been considered in literature.
They correspond to different choices of the remaining two parameters.
Larkin and Ovchinnikov introduced the parameterization\cite{larkin68,Larkin86}
\ber
\label{hh}
\xa = -\ya =h: 
\XX = \mat h & 0 \\ 0 & -\tilde h \matend , \YY = \mat -h & 0 \\ 0 & \tilde h \matend.
\eer
Shelankov's distribution functions \cite{shelankov85} follow from
\ber
\label{FF}
\xa = \yb =F: 
\XX = \mat F & 0 \\ 0 & F \matend , \YY = \mat \tilde F & 0 \\ 0 & \tilde F \matend.
\eer
The author introduced the parameterization \cite{eschrig97}
\ber
\label{xx}
\ya = \yb =0: 
\XX = \mat \xa & 0 \\ 0 & 0 \matend , \YY = \mat 0 & 0 \\ 0 & \xb \matend.
\eer
The advantage of \eqref{xx} is that the transport equations take their simplest
form. The advantage of \eqref{FF} is that $\XX$ and $\YY$ are scalar in 
particle-hole space. And the advantage of \eqref{hh} is that $\XX+\YY=0$.
Why the latter property is an advantage one can see when re-writing
Eq.~\eqref{gqk2} into
\begin{widetext}
\beq
\label{gqk3}
\gqk = -\frac{i\pi }{2}\left[ 
(\XX+\YY) + \frac{\gqr}{-i\pi } \qt (\XX-\YY) - (\XX-\YY) \qt
\frac{\gqa}{-i\pi} - 
\frac{\gqr}{-i\pi} \qt (\XX + \YY) \qt
\frac{\gqa}{-i\pi}  \right]\; .
\eeq
\end{widetext}
With $\XX+\YY=0$ this leads to
\beq
\label{gqk4}
\gqk = \gqr\qt \XX - \XX \qt \gqa \quad \mbox{with}  \quad
\XX = \mat h & 0 \\ 0 & -\tilde h \matend ,
\eeq
which is an equivalent definition to Eq.~\eqref{hh} that was first given by
Larkin and Ovchinnikov.

The symmetry relations for all these distribution functions are 
\ber
\tilde h(\ep ,\pf,\R,t)&=&h(-\ep, -\pf,\R,t)^{\ast }, \\
\Fb(\ep ,\pf,\R,t)&=&\Fa(-\ep,-\pf,\R,t)^{\ast }, \\
\xb(\ep,\pf,\R,t)&=&\xa(-\ep,-\pf,\R,t)^{\ast }, 
\eer
and
\ber
h(\ep,\pf,\R,t) &=&h(\ep,\pf,\R,t)^{\dagger },\\
\Fa(\ep,\pf,\R,t) &=&\Fa(\ep,\pf,\R,t)^{\dagger },\\
\xa(\ep,\pf,\R,t) &=&\xa(\ep,\pf,\R,t)^{\dagger }.
\eer

The $\xa $ and $\xb $ are expressed in terms of the other
distribution functions in a straightforward way, and we obtain
Eqs.~\eqref{xaF}-\eqref{xtoh} of the main text.

Finally we comment on the linear response, Eq.~\eqref{dXXYYra}. 
Here, the most convenient parameterization is
\beq
\label{dXXra}
\dWWra = \mat 0 & \dgara \\ 0 & 0 \matend , \quad
\dZZra = \mat 0 & 0 \\ \dgbra & 0 \matend \; . 
\eeq

\section{Properties of the equations of motion}
\label{properties}
In this Appendix, we use some shorthand notation
in order not to be confused by too cumbersome expressions.
\renewcommand{\ra}{\scriptscriptstyle \rm X}
We use for the superscripts ($R$, $A$, $M$) the notation ($X$).
We parameterize the position on the trajectory by a spatial coordinate
$\vec{R}=\rho \vf $.
We also introduce the symbol $\partial $ for $\hbar \qpartial $
and omit the $\qt $ symbol in all products.
Finally, we use the shorthand notation
$\Eara = \epsilon - \vara $, $\Ebra = -\epsilon - \vbra$,
and $\Eak = -\vak $, $\Ebk = - \vbk$.
\renewcommand{\qt}{}
\subsection{Relations between different solutions for \state functions}
\label{relations}
We consider solutions of the equations of motion for the \state functions, Eq.~\ref{cricc},
and for simplicity we concentrate on the first one, as the second is
related to the first by fundamental symmetry relations. 
The equation,
\ber
\label{riccatti}
i\partial \gara - \gara \qt \Dbra \qt \gara + 
\Eara \qt \gara - \gara \qt \Ebra  + \Dara &=&0 , \nonumber \\
\gara(0) &=& \gara_i,\qquad
\eer
is a Riccati matrix differential equation, the basic properties of which
were thoroughly studied, e.g. in the book of Reid.\cite{Reid}
Associated with any solution $\gara (\rho)$ of (\ref{riccatti}) 
are three quantities $\gaqra (\rho|\gara)$, $\haqra (\rho|\gara )$, and
$\faqra (\rho|\gara)$, which obey the set of equations
\ber
\label{ghf1}
i\partial \gaqra + (\Eara-\gara \qt \Dbra)\qt \gaqra &=& 0,
\qquad \gaqra (0)=1 \; ,\qquad\\
\label{ghf2}
i\partial \haqra - \haqra \qt (\Ebra+\Dbra \qt \gara)&=& 0,
\qquad \haqra (0)=1 \; ,\\
\label{ghf3}
i\partial \faqra + \haqra \qt \Dbra \qt \gaqra &=& 0,
\qquad \faqra (0)=0 \; .
\eer

Let us assume we know the solution $\gara_0(\rho) $ with 
initial condition $\gara_0(0)=\gara_{0i}$ and associated functions
$\gaqra_0 $, $\haqra_0$, and $\faqra_0 $. Often it is the case that
we have boundary conditions, which have to be fulfilled for given
molecular fields, external fields, and order parameters. Then we have
to find the initial value $\gara_{0i}$ self consistently.
A property of Riccati differential equations is that the knowledge of one
solution allows to construct any other solution.
For this we note that the solutions $\gaqra(\rho)$, $\haqra(\rho)$, and $\faqra(\rho)$ for
any other initial condition $\gara_i=\gara_{0i}+ \delta^{\ra}_i $ are
\ber
\label{trafo1}
\gaqra(\rho) &=& \gaqra_0(\rho) \qt [1+\delta^{\ra}_i \qt \faqra_0 (\rho) ]^{-1}  \; ,\nonumber \\
\haqra(\rho) &=& [1+\faqra_0 (\rho) \qt \delta^{\ra}_i ]^{-1} \qt \haqra_0(\rho)  \; ,\nonumber \\
\faqra(\rho) &=& [1+\faqra_0 (\rho) \qt \delta^{\ra}_i ]^{-1} \qt \faqra_0(\rho) \; , \nonumber \\
&=& \faqra_0(\rho) \qt [1+\delta^{\ra}_i \qt \faqra_0 (\rho)]^{-1}.
\eer
The full solution $\gara (\rho) $ along the entire trajectory for the new initial
condition is then obtained by the following formula:
\ber
\label{trafo2}
\gara(\rho)&=& \gara_0(\rho)+\gaqra_0(\rho)\qt \delta^{\ra}_i \qt \haqra (\rho) \; \nonumber \\
&=& \gara_0(\rho)+ \gaqra (\rho) \qt \delta^{\ra}_i \qt \haqra_0(\rho)  . 
\eer

\subsection{Integral equation for coherence amplitudes}
For the retarded and advanced coherence amplitudes there is a possibility
to formulate the Riccati differential equation
as an integral equation.
The formal solutions of equations (\ref{ghf1}-\ref{ghf3}) are
\ber
\gaqra (\rho)&=& {\cal P} e^{i\int_{0}^\rho (\Eara-\gara \qt \Dbra ) d\rho''} \nonumber \\
\haqra (\rho)&=& \overline{\cal P} e^{-i\int_{0}^\rho (\Ebra+\Dbra \qt \gara ){\rm d}\rho''} \\
\gaqra (\rho)^{-1}&=& \overline{\cal P} e^{-i\int_{0}^\rho (\Eara-\gara \qt \Dbra ) d\rho''} \nonumber \\
\haqra (\rho)^{-1}&=& {\cal P} e^{i\int_{0}^\rho (\Ebra+\Dbra \qt \gara ){\rm d}\rho''}
\eer
where ${\cal P} $ ($\overline{\cal P}$)is a trajectory (anti-) path-ordering operator.
With the definition of the transfer operators
\ber
S^X_U(\rho,\rho')=
\gaqra (\rho)\gaqra (\rho')^{-1}&=& {\cal P} e^{i\int_{\rho'}^\rho (\Eara-\gara \qt \Dbra ) {\rm d}\rho''} 
\nonumber \\
S^X_V(\rho',\rho)=
\haqra (\rho')^{-1}\haqra (\rho)&=& 
{\cal P} e^{i\int_{\rho}^{\rho'} (\Ebra+\Dbra \qt \gara ){\rm d}\rho''}
\nonumber \\
\label{transf}
\eer
and introducing the notation
\ber
I^{\ra}_{\slDelta } (\rho)=
-\Dara (\rho) - \gara (\rho)\qt \Dbra (\rho)\qt \gara (\rho)
\eer
we can write the equation of motion as
\ber
&&i\partial \; \gara + ( \Ea -\ga \qt \Db )^{\ra} \qt \gara -
\gara \qt ( \Eb +\Db \qt \ga )^{\ra} = I_{\slDelta }^{\ra } \qquad
\eer
and obtain an integral equation for $\gara $,
\ber
\gara (\rho)&=& S^X_U(\rho,0) \gara (0) S^X_V(0,\rho) \nonumber \\
&-&i \int_{0}^\rho S^X_U(\rho,\rho') 
\qt I^{\ra }_{\slDelta } (\rho') \qt
S^X_V(\rho',\rho) \; {\rm d}\rho'  \; .
\eer

\subsection{Construction of solutions for distribution functions}
In a similar way we can obtain integral representations for the Keldysh
Green's functions.
Consider the transport equation for the distribution function $\xa $,
\ber
&&i\partial \xa + ( \Ea -\ga \qt \Db )^{\ret } \qt \xa -
\xa \qt ( \Ea +\Da \qt \gb )^{\adv } =  I^{\kel } \nonumber \\
&&I^{\kel} = \gar \qt \Ebk \qt \gba +\Dak \qt \gba + \gar \qt \Dbk +\Eak .
\eer
The solutions can be written in terms of 
$S^{\ret}_U(\rho,\rho')$ and $\tilde S^{\adv }_V(\rho,\rho')$, Eq.~\eqref{transf},  as
\ber
\xa (\rho)&= &
S_U^{\ret }(\rho,0) \qt
\xa (0) 
\tilde S_V^{\adv }(0,\rho)
\nonumber \\
&-&i \int_0^\rho 
S_U^{\ret }(\rho,\rho') \qt
I^\kel(\rho') \qt
\tilde S_V^{\adv }(\rho',\rho)
{\rm d}\rho' 
\qquad
\eer

\subsection{Construction of solutions for linear response functions}
Analogously we obtain the linear response equations for retarded and
advanced coherence functions, that are given by the solutions of
\ber
&&i\partial \; \delta \gara + ( \Ea -\ga \qt \Db )^{\ra} \qt \delta \gara -
\delta \gara \qt ( \Eb +\Db \qt \ga )^{\ra} = \de I^{\ra }\nonumber \\
&&\quad \de I^{\ra }=\gara \qt \delta \Dbra \qt \gara
-\delta \Eara \qt \gara + \gara \qt \delta \Ebra - \delta \Dara .\qquad 
\eer
Its solutions can be written in terms of 
$S^X_U(\rho,\rho')$ and $S^X_V(\rho,\rho')$, Eq.~\eqref{transf},  as
\ber
\dgara (\rho)&= &
S_U^{\ra }(\rho,0) \qt
\dgara (0) \qt
S_V^{\ra }(0,\rho)  \nonumber \\
&-&i \int_0^\rho 
S_U^{\ra }(\rho,\rho') \qt
\de I^{\ra} (\rho')  \qt
S_V^{\ra }(\rho',\rho) 
{\rm d}\rho' 
\; . \qquad
\eer

\renewcommand{\qt }{\circ}
\section{Generalized gauge transformations}
\label{gauge}
We start with the set of quasiclassical equations
\begin{eqnarray}
[\check \epsilon -\check h,\check g]_{\qt } + i\partial \check g  &=& \check 0 \qquad
\qquad \check g \qt \check g = - \pi^2 \check 1 \; .
\label{eq1}
\end{eqnarray}
We note that the generalized gauge transformation in combined Keldysh and
Nambu-Gor'kov space
\begin{eqnarray}
\check g' = \check T ^{-1} \qt \check g \qt \check T
\end{eqnarray}
leaves equations (\ref{eq1}) invariant,
\begin{eqnarray}
[\check \epsilon -\check h',\check g']_{\qt} + i\partial \check g'  &=& \check 0 \quad ,
\quad \check g' \qt \check g' = - \pi^2 \check 1 \; . \qquad
\label{eq2}
\end{eqnarray}
if we use as gauge transformed source terms
\begin{eqnarray}
\check \epsilon-\check h' = \check T ^{-1}\qt
(\check \epsilon- \check h ) \qt \check T + \check T ^{-1} \qt i\partial \check T \; .
\end{eqnarray}
Here, the matrix $\check T$ is of the following form
\begin{equation}
\check T = \left(
\begin{array}{cc}
\hat T^{\ret} & \hat T^{\kel} \\ 0&\hat T^{\adv}
\end{array}
\right)
\end{equation}
We write now
\begin{equation}
\check T = \check T_D + \check T_K =\left(
\begin{array}{cc}
\hat T^{\ret} & 0 \\ 0&\hat T^{\adv}
\end{array}
\right)+
\left(
\begin{array}{cc}
0& \hat T^{\kel} \\ 0&0
\end{array}
\right)
\end{equation}
Here, the matrix $\check T_D$ is assumed to have an inverse $\check T_D^{-1}$.
Then, the inverse of $\check T$ is expressed through $\check T_D^{-1}$ by
$\check T^{-1} = \check T_D^{-1} -  \check T_D^{-1} \qt \check T_K \qt \check T_D^{-1}$.
Defining $\check T_K= -\check T_D \qt \check F$ we can write without loss of
generality
\begin{equation}
\check T = \check T_D \qt \left( \check 1 -\check F \right) 
\qquad
\check T^{-1} = \left( \check 1 +\check F \right) \qt \check T_D^{-1}
\end{equation}
where the matrix structure of $\check F$ is given by
\begin{equation}
\check F =
\left(
\begin{array}{cc}
0&  \hat F \\ 0&0
\end{array}
\right),
\end{equation}
and $\check F\qt \check F=0 $ ensures the simple structure of the inverse
of $\check T$.
Now we can write the generalized gauge transformation as 
\begin{eqnarray}
\check g' = (\check 1 + \check F) \qt \check T_D^{-1} \qt \check g \qt \check T_D
\qt (\check 1 - \check F)
\end{eqnarray}
Thus, we have two types of transformation, which we can study separately,
first for $\check F=\check 0$,
\begin{eqnarray}
\check g' &=& \check T_D^{-1} \qt \check g \qt \check T_D
\nonumber \\
\check \epsilon-\check h' &=& \check T_D^{-1} \qt
(\check \epsilon- \check h ) \qt \check T_D + \check T_D ^{-1} \qt i\partial \check T_D \; .
\end{eqnarray}
and second for $\check T_D = \check 1$,
\begin{eqnarray}
\check g'& =& (\check 1 + \check F) \qt \check g \qt (\check 1 - \check F) =
\check g - [\check g,\check F]_{\qt } \; ,
\nonumber \\
\check \epsilon-\check h' &= &
(\check \epsilon- \check h ) -
[\check \epsilon- \check h ,\check F]_{\qt}
-  i\partial \check F \; .
\end{eqnarray}
The second transformation does not affect the retarded and advanced components,
and redefines only the Keldysh components. It leads to a 
gauge transformation for the distribution functions.
A general transformation is obtained by successive application of these
two types of transformations.

For an infinitesimal transformation $\check T_D = \check 1 - \delta \check T_D$
we obtain to first order
\begin{eqnarray}
\check g'& =& (\check 1 + \delta \check T_D)  \qt
\check g \qt (\check 1 - \delta \check T_D) =
\check g - [\check g,\delta \check T_D]_{\qt } \; ,
\nonumber \\
\check \epsilon-\check h' &= &
(\check \epsilon- \check h ) -
[\check \epsilon- \check h ,\delta \check T_D]_{\qt}
-  i\partial \delta \check T_D \; .
\end{eqnarray}
Note the similarity to the second type of gauge transformation. This follows from
the fact that formally we can write $(\check 1 + \check F)= e^{\check F} $,
and $(\check 1 - \check F)= e^{-\check F} $ due to $\check F \qt \check F =0$, so
the same equations like for $\check T_D = e^{-\delta \check T_D}\approx
\check 1 - \delta \check T_D$ hold.

\subsection{Transformations of \state functions}
The Riccati differential equations are invariant under the following
transformation with transformation matrices $\SRA$ and $\TRA$,\\
\ber
\gara_0&=&\SRAi \qt \gara \qt \TRA, \\
\Dara_0&=&\SRAi \qt \Dara \qt \TRA,\\
\Eara_0&=&\SRAi \qt (i\partial \SRA + \Eara \qt \SRA),\\[0.5cm]
\gaqra_0&=&\SRAi \qt \gaqra \qt\SRA,\\
\haqra_0&=&\TRAi \qt \haqra \qt \TRA,\\
\faqra_0&=&\TRAi \qt \faqra \qt \SRA,\\[0.5cm]
\xa_0&=& \SRi \qt \xa \qt \SA,\\
\Dak_0&=&\SRi \qt \Dak \qt \TA,\\
\Eak_0&=&\SRi \qt \Eak \qt \SA, 
\eer
with $\Eara = \epsilon - \vara $, $\Eak = -\vak $,
and analogous relations for the particle-hole conjugated
quantities.
An important case is that of 
unitary transformation matrices, where
this transformation is a {\it local gauge transformation}, 
possibly accompanied by a {\it local spin rotation}. 
In this case it is more convenient to write
\ber
\TRA = e^{\frac{i}{2}\phi } \qquad \SRA = e^{-\frac{i}{2}\tilde \phi } \; .
\eer
The important feature is the occurrence of the new driving terms
$\SRAi \qt i\partial \SRA $ that gives a contribution 
$\vec{A}_\phi$ to the vector potential.
For gauge transformations they are equal to
\ber
-\frac{e}{c} \vec{v}_f \vec{A}_\phi=
e^{\frac{i}{2}\tilde \phi } \qt 
\left( \frac{\hbar}{2}\; \qpartial \tilde \phi \right) 
\qt e^{-\frac{i}{2}\tilde \phi }
.
\eer
When $\qpartial \tilde \phi$  commutes with $\tilde \phi$,
the two gauge factors on either side cancel in equilibrium.
As can be seen above there is a very broad class of transformations
(not necessarily gauge transformations)
which leave the equations of motion invariant.

\subsection{Transformations of distribution functions}
\label{gauge2}
The equations of motion are also invariant under the transformations \\
\ber
\xa_0&=&\xa - (\Fa_0 + \gar \qt \Fb_0 \qt \gba )\\
\Dak_0&=&\Dak + (\Dar \qt \Fb_0 + \Fa_0 \qt \Daa )\\
\Eak_0&=&\Eak - (\Ear \qt \Fa_0 - \Fa_0 \qt \Eaa )-i\partial \Fa_0
\eer
with $\Eara= -\vara$, $\Eak = -\vak $,
and analogously for $\xb_0$, $\Dbk_0$, and $\Ebk_0$.
A natural choice is the equilibrium distribution function,
$\Fa_0= \tanh (\epsilon/2T)$ (and $\tilde F_0$ related by
symmetry).
The transformed quantities are called
{\it anomalous} in this case.
\\
Let us assume we calculate the Keldysh Green's function from
$\xa $ and $\xb$
and obtain $\hat g^{\kel}[\xa, \xb]$. 
Applying the above transformation of the driving terms 
we could also solve for the $\xa_0 $ and $\xb_0$ instead.
We can then construct an anomalous Green's
function defined by $\hat g^\ano \equiv \hat g^{\kel}[\xa_0, \xb_0]$.
The difference between the Keldysh part and the anomalous part
of the Green's function is called {\it spectral } part of the
Green's function.
If we introduce
\begin{equation}
\hat F_0=
\left(
\begin{array}{cc}
F_0 & 0 \\ 0 & -\tilde F_0
\end{array}
\right)
\end{equation}
then it is given by,
\ber
\hat g^{\kel}[\xa, \xb] -\hat g^{\kel}[\xa_0, \xb_0] = \hat g^{\ret} \qt \hat F_0 - \hat F_0 \hat g^{\adv}
\eer
Thus, it is enough to solve for $\xa_0 $ and $\xb_0$ to obtain directly
the full Keldysh Green's functions once one has the retarded and advanced ones.
The choice of the distribution function $\hat F_0$ is of course somewhat
arbitrary, but it is best chosen to be the equilibrium distribution function
whenever there is one well defined. For a spatially varying 
electrochemical potential $\slPhi(\vec{R})$ and possibly 
varying temperature, 
\ber
\Fa_0= \tanh \left(\frac{\epsilon-e\slPhi(\vec{R})}{2k_{\rm B}T(\vec{R})} \right), 
\tilde \Fa_0= -\tanh \left(\frac{\epsilon+e\slPhi(\vec{R})}{2k_{\rm B}T(\vec{R})} \right), 
\nonumber
\eer
where $\slPhi(\vec{R})$ is
determined by the unit trace of the Keldysh Green's function to ensure
local charge neutrality.
The advantage of such a choice is that the anomalous functions
$\xa_0 $ are zero in `reservoir' regions.
If the electrochemical potentials are different on the two sides of an
interface, then the boundary
conditions produces a nonzero anomalous component
$\xa_0 $ on either side of the interface. 
It is always numerically
advisable to use the $\xa_0 $ with the spectral part subtracted instead
using the full $\xa $. This makes the driving forces explicit and
avoids cancellations between large terms.

Let us finally mention the driving terms for the above choice of 
equilibrium function. They are given in the equation for $\xa_0$
by $-i\partial F_0$, with
\beq
\partial F_0= 
\vec{v}_f \left[ e\vec{E}(\vec{R}) - \grad \mu (\vec{R})
-\frac{\epsilon - e\slPhi(\vec{R})}{T(\vec{R})} \grad T(\vec{R}) \right]
\hbar \partial_{\epsilon } F_0  
\eeq
where $\vec{E}$ is the electric field. This corresponds to the force term
in a Boltzmann equation.
There are additional terms for time dependent forces. 
For instance the term $\epsilon \qt F_0-F_0 \qt \epsilon $ 
is equal to $i\hbar \partial_t F_0$.
Note also the term
$-\Dar \qt \tilde F_0 - F_0 \qt \Daa $ which gives for energy independent gap
as off-diagonal force
\ber
\slDelta \cdot \left( 
\tanh \left(\frac{\epsilon+e\slPhi(\vec{R})}{2k_{\rm B}T(\vec{R})}\right) -
\tanh \left(\frac{\epsilon-e\slPhi(\vec{R})}{2k_{\rm B}T(\vec{R})}\right) 
\right).
\eer
Finally, we mention the possibility to define spin dependent forces in a
similar way.

\renewcommand{\ra}{{\scriptscriptstyle \rm R,A}}
\section{Retarded-advanced symmetries and Keldysh symmetries}
\label{symm}
The following symmetries connect retarded and advanced functions and
express symmetries in the Keldysh components:
\ber
\gaa &=& (\gbr )^{\dagger }, \; \,
\Daa = -(\Dbr )^{\dagger },\; \;
\Eaa = (\Ear )^{\dagger },\qquad\\
\gaqa&=& (\hbqr )^{\dagger },\;
\haqa= \plus (\gbqr )^{\dagger },\;
\faqa= \plus (\fbqr )^{\dagger },\\
\xa &=& (\xa )^{\dagger },\; \; \;\;
\Dak = \plus (\Dbk )^{\dagger },\;
\Eak = -(\Eak )^{\dagger } 
\eer
with $\Eara = \epsilon - \vara $, $\Eak = -\vak $.
The quantities $\gaqra $, $\haqra$, and $\faqra$ are defined in Eqs.~\eqref{trafo1}.
Analogous relations hold for the particle-hole conjugated
quantities.

\section{Full solutions in superconductor for S/HM interface}
\label{Full}
The solutions of the boundary conditions 
for the model discussed in section \ref{SHcoh} can be obtained also
in the superconductor explicitely. 
With the abbreviations $\vartheta_{\uparrow \uparrow} = (\vartheta_u-\vartheta_v)/4$,
$P=\sin \frac{\vartheta}{2}\sin \alpha/(1+r) $ and $Q=\cos \frac{\vartheta}{2}\sin \alpha /(1+r)$, 
they are given by
\begin{widetext}
\begin{eqnarray}
(\slGamma_{\rm S})_{(\uparrow\downarrow -\downarrow\uparrow )/2 }&=& \frac{1}{2}\;
\frac{-i Pt^2 (1+r) [\tilde \gamma_{\rm F} \gamma_{\rm S}^2 e^{-i\phi}+\gamma_{\rm F}e^{i\phi}]
+\gamma_{\rm S} (1-\gamma_{\rm F}\tilde \gamma_{\rm F}) \left[ 2r \cos \vartheta -P^2t^4 \right]
}{1-r^2\gamma_{\rm F}\tilde\gamma_{\rm F} + i Pt^2 (1+r) \tilde \gamma_{\rm F} \gamma_{\rm S} e^{-i\phi} }\\
(\slGamma_{\rm S})_{(\uparrow\downarrow +\downarrow\uparrow )/2}&=& \frac{1}{2}\;
\frac{-Qt^2 (1+r) [\tilde \gamma_{\rm F} \gamma_{\rm S}^2 e^{-i\phi}+\gamma_{\rm F}e^{i\phi}]
+i\gamma_{\rm S} (1-\gamma_{\rm F}\tilde \gamma_{\rm F}) \left[ 2r \sin \vartheta +
PQt^4 \right]
}{1-r^2\gamma_{\rm F}\tilde\gamma_{\rm F} + i Pt^2 (1+r) \tilde \gamma_{\rm F} \gamma_{\rm S} e^{-i\phi} }
\end{eqnarray}
\begin{eqnarray}
(\slGamma_{\rm S})_{\uparrow\uparrow }&=& e^{2i\vartheta_{\uparrow\uparrow }}e^{-i\phi }\;
\frac{-t^2 \left[ \tilde \gamma_{\rm F} \gamma_{\rm S}^2 e^{-i\phi}\sin^2 \frac{\alpha}{2} -\gamma_{\rm F} e^{i\phi} \cos^2 \frac{\alpha}{2} \right]
+iPt^2\gamma_{\rm S} \left[ (1+r\gamma_{\rm F}\tilde \gamma_{\rm F}) \sin^2 \frac{\alpha}{2}
+ (r+\gamma_{\rm F}\tilde \gamma_{\rm F}) \cos^2 \frac{\alpha}{2} \right]
}{1-r^2\gamma_{\rm F}\tilde\gamma_{\rm F} + i Pt^2 (1+r) \tilde \gamma_{\rm F} \gamma_{\rm S} e^{-i\phi} }
\quad \quad \\
(\slGamma_{\rm S})_{\downarrow\downarrow }&=& e^{-2i\vartheta_{\uparrow\uparrow }}e^{i\phi }\;
\frac{-t^2 \left[ \tilde \gamma_{\rm F} \gamma_{\rm S}^2 e^{-i\phi }\cos^2 \frac{\alpha}{2} -\gamma_{\rm F}e^{i\phi } \sin^2 \frac{\alpha}{2} \right]
+iPt^2\gamma_{\rm S} \left[ (1+r\gamma_{\rm F}\tilde \gamma_{\rm F}) \cos^2 \frac{\alpha}{2} 
+ (r+\gamma_{\rm F}\tilde \gamma_{\rm F}) \sin^2 \frac{\alpha}{2} \right]
}{1-r^2\gamma_{\rm F}\tilde\gamma_{\rm F} + i Pt^2 (1+r) \tilde \gamma_{\rm F} \gamma_{\rm S} e^{-i\phi} }.
\end{eqnarray}
\end{widetext}
For zero misalignment angle $\alpha$, these solutions simplify: $(\slGamma_{\rm S})_{(\uparrow\downarrow -\downarrow\uparrow )/2 }=
\gamma_{\rm S}  r \cos (\vartheta )
(1-\gamma_{\rm F}\tilde \gamma_{\rm F} )/(1-r^2\gamma_{\rm F}\tilde\gamma_{\rm F} )$,
$(\slGamma_{\rm S})_{(\uparrow\downarrow +\downarrow\uparrow )/2}=
i\gamma_{\rm S} r \sin (\vartheta )
(1-\gamma_{\rm F}\tilde \gamma_{\rm F} )/(1-r^2\gamma_{\rm F}\tilde\gamma_{\rm F})$,
$(\slGamma_{\rm S})_{\uparrow\uparrow }=  t^2 
\gamma_{\rm F} e^{2i\vartheta_{\uparrow\uparrow }}/(
1-r^2\gamma_{\rm F}\tilde\gamma_{\rm F} )$, and
$(\slGamma_{\rm S})_{\downarrow\downarrow }=
-t^2  \tilde \gamma_{\rm F} \gamma_{\rm S}^2 e^{-2i\vartheta_{\uparrow\uparrow }}/(
1-r^2\gamma_{\rm F}\tilde\gamma_{\rm F} )$.

\end{document}

%% file: macros.tex
\newcommand{\bra}[1]{\langle {#1}  \! \mid}
\newcommand{\ket}[1]{\mid  \!  {#1} \rangle}

\newcommand{\half}{\frac{1}{2}}

\newcommand{\sign}{\mbox{sign}}
\newcommand{\plus}{\; \; \, }
\newcommand{\mbf}[1]{\mbox{ \boldmath $ #1$}}

\newcommand{\pf}{{\bfm p}_{\rm F}}

\newcommand{\vf}{{\bfm v}_{\rm F}}

\renewcommand{\vec}[1]{{\bfm{#1}}}

\newcommand{\beq} {\begin{equation}}
\newcommand{\eeq} {\end{equation}}
\newcommand{\ber} {\begin{eqnarray}}
\newcommand{\eer} {\end{eqnarray}}
\newcommand{\nn}{\nonumber }

\newcommand{\de}   {{\delta}}

\newcommand{\epp}[2] {\mbox{$#1\!+\!{#2\over 2}$}}
\newcommand{\epm}[2] {\mbox{$#1\!-\!{#2\over 2}$}}

\newcommand{\R} {{\bfm R}}
\newcommand{\name}[1] {{#1}}

\newcommand{\tc}{\op{\tau }_3 }

\newcommand{\ep}{\epsilon }
\newcommand{\om}{\omega }

\newcommand{\mat}{\left(\begin{array}{cc}}
\newcommand{\matend}{\end{array}\right)}
\newcommand{\matl}{\left(\begin{array}{c}}
\newcommand{\matlend}{\end{array}\right)}
\newcommand{\qt}{{\otimes}}

\newcommand{\op}[1]{\hat{#1}}

\newcommand{{\ra}}{{\scriptscriptstyle \rm R,A}}
\newcommand{{\rak}}{{\scriptscriptstyle \rm R,A,K}}
\newcommand{{\ret}}{{\scriptscriptstyle \rm R}}
\newcommand{{\adv}}{{\scriptscriptstyle \rm A}}
\newcommand{\kel}{{\scriptscriptstyle \rm K}}

\newcommand{\ano}{{\scriptstyle \rm a}}

\newcommand{\PPp}{\check{P}_{+}}
\newcommand{\PPm}{\check{P}_{-}}
\newcommand{\PPpm}{\check{P}_{\pm}}

\newcommand{\Pa}{\op{P}_{+}}
\newcommand{\Pb}{\op{P}_{-}}
\newcommand{\Par}{\Pa^{\ret}}
\newcommand{\Pbr}{\Pb^{\ret}}
\newcommand{\Paa}{\Pa^{\adv}}
\newcommand{\Pba}{\Pb^{\adv}}
\newcommand{\Para}{\Pa^{\ra}}
\newcommand{\Pbra}{\Pb^{\ra}}

\newcommand{\gq}{\op{g}}
\newcommand{\gqr}{\gq^{\ret}}
\newcommand{\gqa}{\gq^{\adv}}
\newcommand{\gqra}{\gq^{\ra}}
\newcommand{\gqk}{\gq^{\kel}}

\newcommand{\dgqra}{\de \! \gqra}

\newcommand{\gaq}{U}
\newcommand{\gbq}{\tilde{\gaq}}

\newcommand{\gbqr}{\gbq^{\ret}}
\newcommand{\gaqa}{\gaq^{\adv}}

\newcommand{\gaqra}{\gaq^{\ra}}

\newcommand{\faq}{W}
\newcommand{\fbq}{\tilde{\faq}}

\newcommand{\fbqr}{\fbq^{\ret}}
\newcommand{\faqa}{\faq^{\adv}}

\newcommand{\faqra}{\faq^{\ra}}

\newcommand{\haq}{V}
\newcommand{\hbq}{\tilde{\haq}}

\newcommand{\hbqr}{\hbq^{\ret}}
\newcommand{\haqa}{\haq^{\adv}}

\newcommand{\haqra}{\haq^{\ra}}

\newcommand{\ga}{\gamma}
\newcommand{\gb}{\tilde{\ga}}
\newcommand{\gar}{\ga^{\ret}}
\newcommand{\gbr}{\gb^{\ret}}
\newcommand{\gaa}{\ga^{\adv}}
\newcommand{\gba}{\gb^{\adv}}
\newcommand{\gara}{\ga^{\ra}}
\newcommand{\gbra}{\gb^{\ra}}
\newcommand{\Ga}{\slGamma}
\newcommand{\Gb}{\sltilde{\Ga}}
\newcommand{\Gar}{\Ga^{\ret}}
\newcommand{\Gbr}{\Gb^{\ret}}
\newcommand{\Gaa}{\Ga^{\adv}}
\newcommand{\Gba}{\Gb^{\adv}}
\newcommand{\Gara}{\Ga^{\ra}}
\newcommand{\Gbra}{\Gb^{\ra}}

\newcommand{\dgara}{\de \! \gara}
\newcommand{\dgbra}{\de \! \gbra}

\newcommand{\xa}{x}
\newcommand{\xb}{\tilde{x}}

\newcommand{\ya}{y}
\newcommand{\yb}{\tilde{y}}

\newcommand{\Xa}{X}
\newcommand{\Xb}{\tilde{X}}

\newcommand{\Fa}{F}
\newcommand{\Fb}{\tilde{\Fa}}

\newcommand{\XX}{\op{X}^{\kel}}
\newcommand{\dZZ}{\de \! \op{Z}}
\newcommand{\dWW}{\de \! \op{W}}
\newcommand{\dWWra}{\de \! \op{W}^{\ra}}
\newcommand{\dZZra}{\de \! \op{Z}^{\ra}}

\newcommand{\YY}{\op{Y}^{\kel}}

\newcommand{\Da}{\slDelta}

\newcommand{\Db}{\sltilde{\Da}}
\newcommand{\Dar}{\Da^{\ret}}
\newcommand{\Dbr}{\Db^{\ret}}
\newcommand{\Daa}{\Da^{\adv}}

\newcommand{\Dara}{\Da^{\ra}}
\newcommand{\Dbra}{\Db^{\ra}}

\newcommand{\Dak}{\Da^{\kel}}
\newcommand{\Dbk}{\Db^{\kel}}

\newcommand{\va}{\slSigma}
\newcommand{\vb}{\sltilde{\va}}
\newcommand{\vak}{\va^{\kel}}
\newcommand{\vbk}{\vb^{\kel}}
\newcommand{\vara}{\va^{\ra}}
\newcommand{\vbra}{\vb^{\ra}}

\newcommand{\Ea}{E}
\newcommand{\Eb}{\tilde{\Ea}}
\newcommand{\Ear}{\Ea^{\ret}}

\newcommand{\Eaa}{\Ea^{\adv}}

\newcommand{\Eara}{\Ea^{\ra}}
\newcommand{\Ebra}{\Eb^{\ra}}
\newcommand{\Eak}{\Ea^{\kel}}
\newcommand{\Ebk}{\Eb^{\kel}}

\newcommand{\hk}{\op{h}^{\kel}}
\newcommand{\hr}{\op{h}^{\ret}}
\newcommand{\ha}{\op{h}^{\adv}}
\newcommand{\hra}{\op{h}^{\ra}}
\newcommand{\gmk}{\op{g}^{\kel}}
\newcommand{\gmr}{\op{g}^{\ret}}
\newcommand{\gma}{\op{g}^{\adv}}

\newcommand{\qpartial}{\vf \!\cdot\! \mbox{\boldmath $\nabla $}}

\newcommand{\grad}{\mbox{\boldmath $\nabla $}}

\newcommand{\ce}{\check{1}}
\newcommand{\cz}{\check{0}}
\newcommand{\cg}{\check{g}}

\newcommand{\TT}{\tilde{T}}

\newcommand{\TA}{T^{\adv}}
\newcommand{\TRA}{T^{\ra}}
\newcommand{\SR}{\TT^{\ret}}
\newcommand{\SA}{\TT^{\adv}}
\newcommand{\SRA}{\TT^{\ra}}

\newcommand{\TRAi}{(\TRA )^{-1}}
\newcommand{\SRi}{(\SR )^{-1}}

\newcommand{\SRAi}{(\SRA )^{-1}}

%% file: Paper.bbl
\begin{thebibliography} {999}
\bibitem{footexp}
Expansion parameters of the theory are {\it not} parameters of the Hamiltonian, 
but the available fraction of the phase space volume 
that is accessible for quasiparticles.
\bibitem{footanom}
This terminology must not be confused with the off-diagonal part
of the Nambu-Gor'kov Green's function, that also often is called ``anomalous''.
\bibitem{footRn}
Note that 
because $\vec{v}_{\rm F}(\vec{p}_{\rm F})/|\vec{v}_{\rm F}(\vec{p}_{\rm F})|=\vec{n}_{\rm F}(\vec{p}_{\rm F})$
is the Fermi surface normal, the geometric relation
$(dp_{\rm F}) \vec{n}_{\rm F}(\vec{p}_{\rm F}) \cdot \vec{e}_\perp=(dp_{||})$ holds.
\end{thebibliography}
